\documentclass[10pt,journal,compsoc]{IEEEtran}
\usepackage{ragged2e}
\usepackage{amsmath}
\usepackage{subfigure}
\usepackage{graphicx}
\usepackage{multirow}
\usepackage{amsthm}
\usepackage[font=bf]{caption}
\usepackage{amsfonts}

\newtheorem{mydef}{Definition}
\DeclareMathOperator*{\argmax}{arg\,max}
\ifCLASSOPTIONcompsoc
  \usepackage[nocompress]{cite}
\else
  \usepackage{cite}
\fi

\ifCLASSINFOpdf
\else
\fi

\begin{document}
%
\title{Attributed Social Network Embedding}

\author{Lizi~Liao,~
        Xiangnan~He,~
        Hanwang~Zhang,~
        and~Tat-Seng~Chua
\IEEEcompsocitemizethanks{
\IEEEcompsocthanksitem X. He is the corresponding author. E-mail: xiangnanhe@gmail.com
\IEEEcompsocthanksitem L. Liao is with the NUS Graduate School for Integrative Sciences and Engineering, National University of Singapore, Singapore, 117456.\protect\\
E-mail: liaolizi.llz@gmail.com
\IEEEcompsocthanksitem X. He, H. Zhang and TS. Chua are with National University of Singapore.}
\thanks{Manuscript received May 12, 2017; revised **** ****.}}

%

\markboth{Journal of \LaTeX\ Class Files,~Vol.~14, No.~8, May~2017}%
{Shell \MakeLowercase{\textit{et al.}}: Bare Demo of IEEEtran.cls for Computer Society Journals}

\IEEEtitleabstractindextext{%
\justify
\begin{abstract}
Embedding network data into a low-dimensional vector space has shown promising performance for many real-world applications, such as node classification and entity retrieval. However, most existing methods focused only on leveraging network structure. For social networks, besides the network structure, there also exists rich information about social actors, such as user profiles of friendship networks and textual content of citation networks. These rich attribute information of social actors reveal the homophily effect, exerting huge impacts on the formation of social networks.
In this paper, we explore the rich evidence source of attributes in social networks to improve network embedding. We propose a generic Social Network Embedding framework (\textit{SNE}), which learns representations for social actors (\textit{i.e.}, nodes) by preserving both the \textit{structural proximity} and \textit{attribute proximity}. While the \textit{structural proximity} captures the global network structure, the \textit{attribute proximity} accounts for the homophily effect. To justify our proposal, we conduct extensive experiments on four real-world social networks. Compared to the state-of-the-art network embedding approaches, \textit{SNE} can learn more informative representations, achieving substantial gains on the tasks of link prediction and node classification. Specifically, \textit{SNE} significantly outperforms \textit{node2vec} with an $8.2\%$ relative improvement on the link prediction task, and a $12.7\%$ gain on the node classification task.  
\end{abstract}

}

\maketitle

\IEEEdisplaynontitleabstractindextext

\IEEEpeerreviewmaketitle

\IEEEraisesectionheading{\section{Introduction}\label{sec:introduction}}
\IEEEPARstart{S}{ocial} networks are an important class of networks that span a wide variety of media, ranging from social websites such as Facebook and Twitter, citation networks of academic papers, and telephone caller--callee networks --- to name a few. Many applications need to mine useful information from social networks. 
For instance, content providers need to cluster users into groups for targeted advertising \cite{wang2017}, and recommender systems need to estimate the preference of a user on items for personalized recommendation \cite{he2017birank}. 
In order to apply general machine learning techniques on network-structured data, it is essential to learn informative node representations. 

Recently, research interest in representation learning has spread from natural language to network data~\cite{perozzi2014deepwalk}.
Many network embedding methods have been proposed~\cite{chang2015heterogeneous,grovernode2vec,perozzi2014deepwalk,tang2015line}, and show promising performance for various applications. However, existing methods primarily focused on general class of networks and leveraged the structural information only.
For social networks, we point out that there almost always exists rich information about social actors in addition to the link structure.
For example, users on social websites may have profiles like age, gender and textual comments. We term all such auxiliary information as \textit{attributes}, which not only refer to user demographics, but also include other information such as the affiliated texts and the possible labels.
\begin{figure}[!htp]
	\centering
	\subfigure[][class year]{\includegraphics[scale = 0.2]{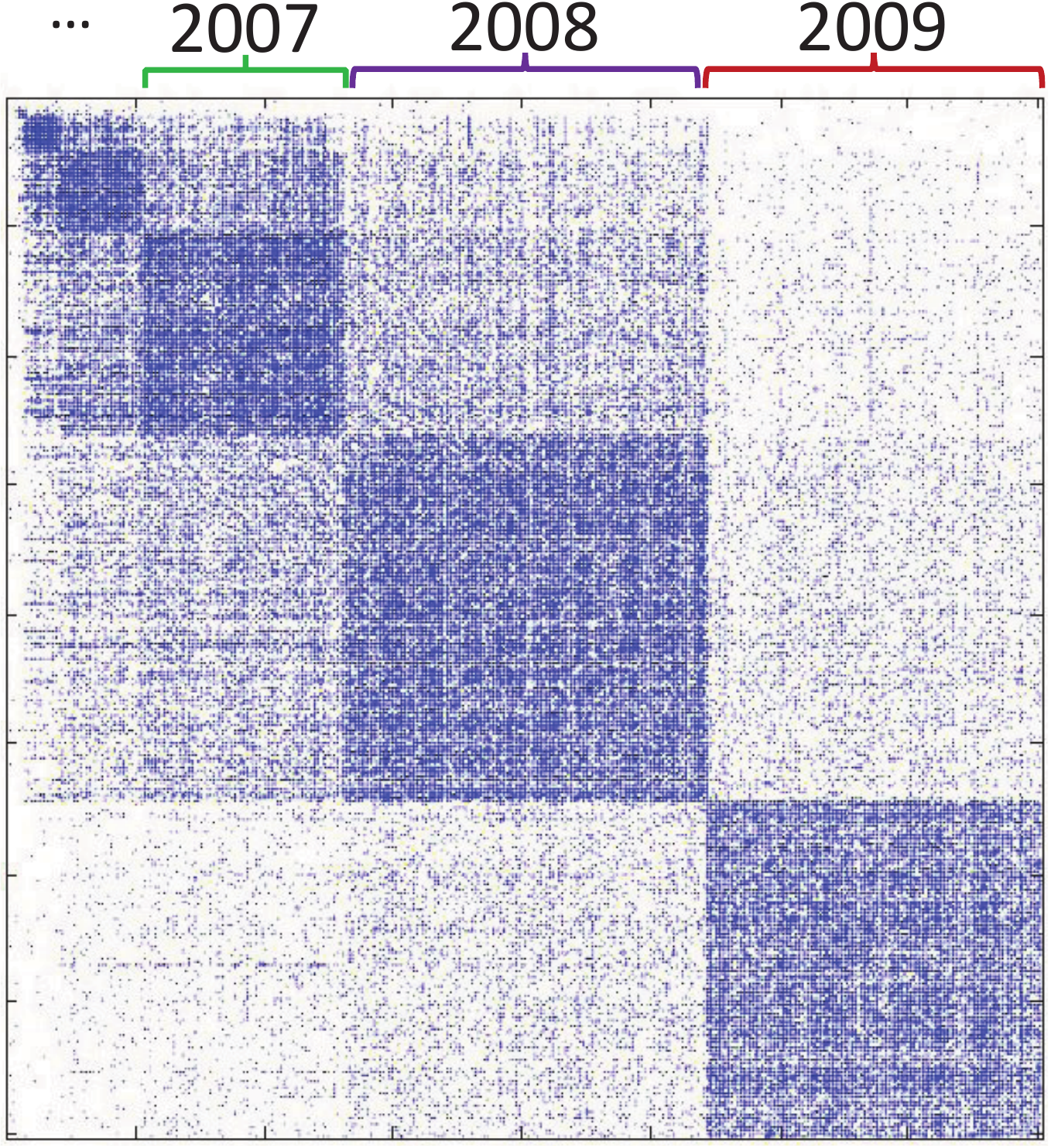}}
	\subfigure[][major]{\includegraphics[scale = 0.2]{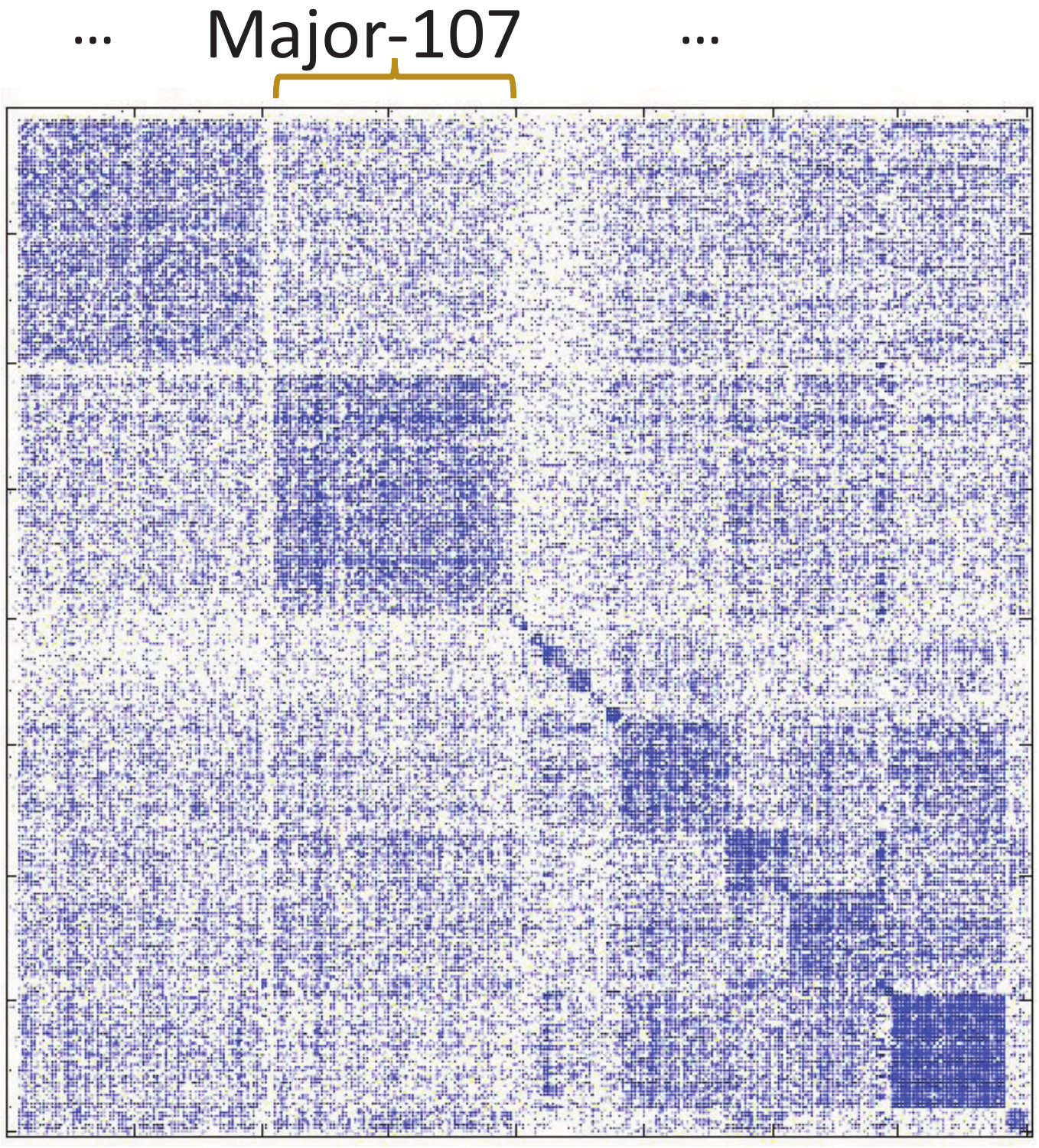}}
	\subfigure[][dormitory]{\includegraphics[scale = 0.2]{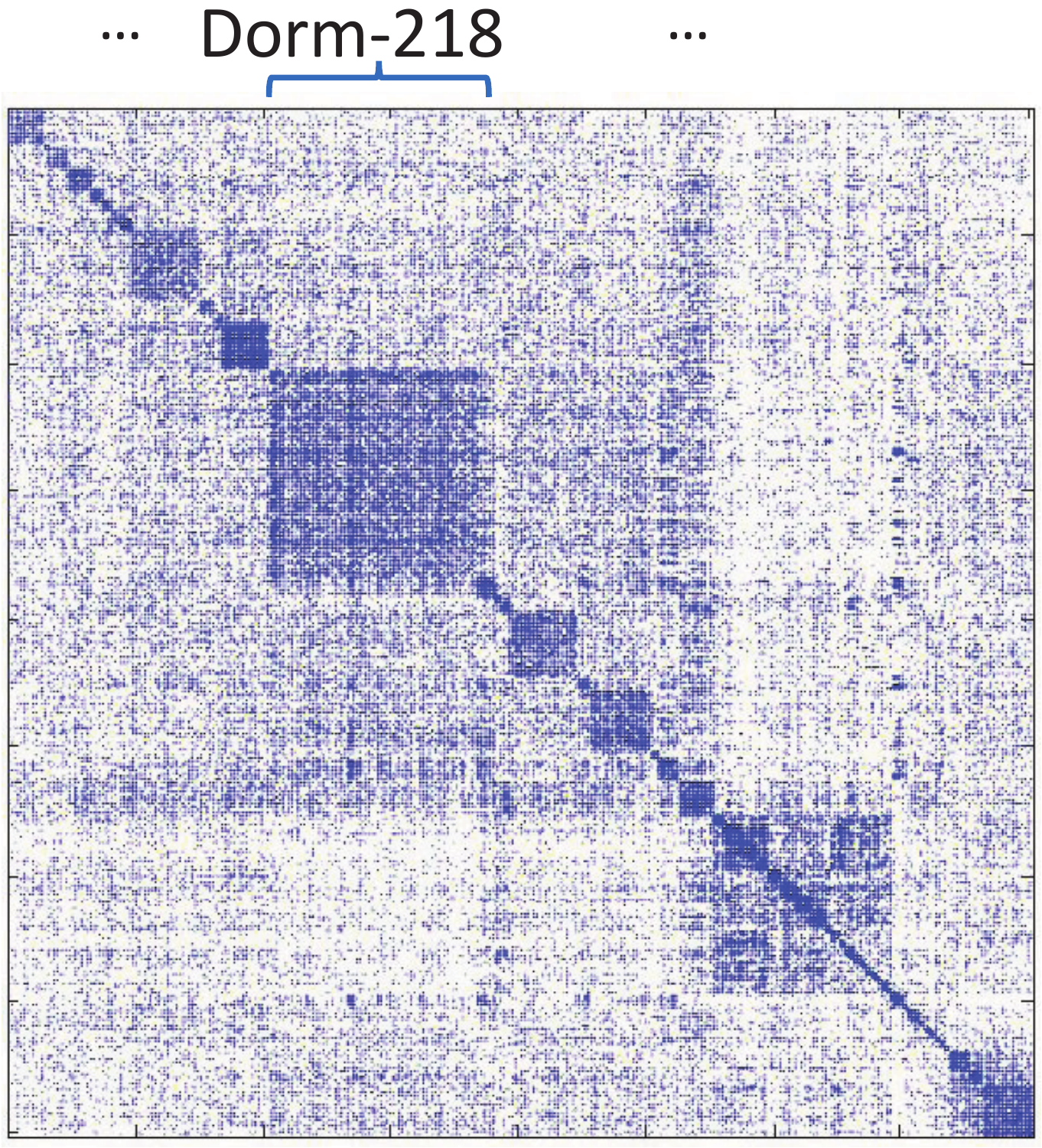}}
	
	\caption{Attribute homophily largely impacts social network: we group users in each 4018$\times$4018 user matrix based on a specific attribute. Clear blocks around the diagonal show the attribute homophily effect.}
	\label{Fig:block}
	\vspace{-0.2cm}
\end{figure}

Attributes essentially exert huge impacts on the organization of social networks. Many studies have justified its importance, ranging from user demographics~\cite{burger2011discriminating}, to subjective preference like political orientation and personal
interests~\cite{pennacchiotti2011democrats}.
To illustrate this point, we plot the user--user friendship matrix of a Facebook dataset from three views\footnote{This is the Chapel Hill data constructed by \cite{traud2012social}, which we will detail later in Section~\ref{ss:datasets}.}. 
Each row or column denotes a user, and a colored point indicates that the corresponding users are friends.
Each subfigure is a re-ordering of users according to a certain attribute such as ``class year', ``major'' and ``dormitory''. For example, Figure~\ref{Fig:block}(a) first groups users by the attribute ``class year'', and then sort these resulting groups in chronological order.
As can be seen, there exist clear block structures in each subfigure, where users of a block are more densely connected. Each block actually points to users of the same attribute; for example, the right bottom block of Figure~\ref{Fig:block}(a) corresponds to users who will graduate in the year of 2009. This real-world example lends support to the importance of attribute homophily. 
By jointly considering the attribute homophily and the network structure, we believe more informative node representations can be learned. 
Moreover, since we utilize the auxiliary attribute information, the link sparsity and cold-start problem~\cite{wangstructural} can largely be alleviated.

In this paper, we present a neural framework named \textit{SNE} for learning node representations from social network data. \textit{SNE} is a generic machine learner working with
real-valued feature vectors, where each feature denotes the ID or an attribute of a node.
Through this, we can easily incorporate any type and number of attributes.
Under our \textit{SNE} framework, each feature is associated with an embedding, and the final embedding for a node is aggregated from its ID embedding (which preserves the \textit{structural proximity}) and attribute embedding (which preserves the \textit{attribute proximity}).
To capture the complex interactions between features, we adopt a multi-layer neural network to take advantage of strong representation and generalization ability of deep learning.

In summary, the contributions of this paper are as follows.
\begin{itemize}
	\item  We demonstrate the importance of integrating network structure and attributes for learning more informative node representations for social networks.
	\item  We propose a generic framework \textit{SNE} to perform social network embedding by preserving the \textit{structural proximity} and \textit{attribute proximity} of social networks. 
	\item  We conduct extensive experiments on four datasets with two tasks of link prediction and node classification. Empirical results and case studies demonstrate the effectiveness and rationality of \textit{SNE}.
\end{itemize}

The rest of the paper is organized as follows. We first discuss the related work in Section 2, followed by providing some preliminaries in Section 3. We then present the \textit{SNE} framework in Section 4. We show experimental results in Section 5, before concluding the whole paper in Section 6.

\section{Related Work}
In this section, we briefly summarize studies about attribute homophily. We then discuss network embedding methods that are closely related to our work.

\subsection{Attribute homophily in Social Networks}
Social networks belong to a special class of networks, in which the formation of social ties involves not only the self-organizing network process, but also the attribute-based process \cite{robins2011exponential}. The motivation for considering \textit{attribute proximity} in the embedding procedure is rooted in the large impact of attribute homophily, which plays an important role in attribute-based process. Therefore, we provide a brief summarization of homophily studies here as a background. Generally speaking, the ``homophily principle''---birds of a feather flock together---is one of the most striking and robust empirical regularities of social life \cite{lazarsfeld1954friendship,laumann1966prestige,mcpherson2001birds}. The hypothesis that people similar to each other tend to become friends dates back to at least the 70s in the last
century. In social science, there is a general expectation
that individuals develop friendships with others of approximately the same age \cite{kurth1970friendships}. In \cite{mcpherson1987homophily} the authors studied the inter-connectedness between homogeneous composition of groups and the emergence of homophily. In \cite{fiore2005homophily} the authors tried to find the role of homophily in online dating choices made by users. They found that online users of the online dating system seek people like them much more often than chance would predict, just as in the offline world. In more recent years, \cite{kossinets2009origins} investigated the origins of homophily in a large university community, using network data in which interactions, attributes and affiliations were all recorded over time. Not surprisingly, it has been concluded that besides structural proximity, preferences for attribute similarity also provides an important factor for the social network formation procedure. Thus, to get more informative representations for social networks, we should take attributes information into consideration.

\subsection{Network Embedding}
Some earlier works such as Local Linear Embedding (LLE)~\cite{roweis2000nonlinear}, IsoMAP~\cite{tenenbaum2000global} and Laplacian Eigenmap~\cite{belkin2001laplacian} first transform data into an affinity graph based on the feature vectors of nodes ( e.g., k-nearest neighbors of nodes) and then embed the graph by solving the leading eigenvectors of the affinity matrix.

Recent works focus more on embedding an existing network into a low-dimensional vector space to facilitate further analysis and achieve better performance than those earlier works. In \cite{perozzi2014deepwalk} the authors deployed truncated random walks on networks to generate node sequences. The generated node sequences are treated as sentences in language models and fed to the Skip-gram model to learn the embeddings. In \cite{grovernode2vec} the authors modified the way of generating node sequences by balancing breadth-first sampling and depth-first sampling, and achieved performance improvements. Instead of performing simulated ``walks'' on the networks, \cite{tang2015line} proposed clear objective functions to preserve the \textit{first-order proximity} and \textit{second-order proximity} of nodes while \cite{wangstructural} introduced deep models with multiple layers of non-linear functions to capture the highly non-linear network structure. However, all these methods only leverage network structure. In social networks, there exists large amount of attribute information. Purely structure-based methods fail to capture such valuable information, thus may result in less informative embeddings. In addition, these methods get affected easily when the link sparsity problem occurs.

Some recent efforts have explored the possibility of integrating contents to learn better representations \cite{huang2017label}. For example, \textit{TADW} \cite{yang2015network} proposed text-associated \textit{DeepWalk} \cite{perozzi2014deepwalk} to incorporate text features into the matrix factorization framework. However, only text attributes can be handled. Being with the same problem, \textit{TriDNR} \cite{shirui2016} proposed to separately learn embeddings from the structure-based \textit{DeepWalk} \cite{perozzi2014deepwalk} and label-fused \textit{Doc2Vec} model \cite{le2014distributed}, the embeddings learned were linearly combined together in an iterative way. Under such a scheme, the knowledge interaction between the two separate models only goes through a series of weighted sum operations and lacks further convergence constrains. On the contrary, our method models the structure proximity and attribute proximity in an end-to-end neural network that does not have such limitations. Also, by incorporating structure and attribute modeling by an early fusion, the two parts only need to complement each other, resulting in sufficient knowledge interactions \cite{cheng2016wide}.

There have also been efforts explored semi-supervised learning for network embedding. \cite{weston2012deep} combined an embedding-based regularizer with a supervised learner to incorporate label information. Instead of imposing regularization, \cite{yang2016icml} used embeddings to predict the context in graph and leveraged label information to build both transductive and inductive formulations. In our framework, label information can also be incorporated in the same way similar to \cite{yang2016icml} when available. We leave this extension as future work, as this work focuses on the modeling of attributes for network embedding.

\section{Definitions}
Social networks are more than links; in most cases, social actors are associated with rich attributes. 
We denote a social network as $G = (\mathcal{U}, \mathcal{E}, \mathcal{A})$, where $\mathcal{U}=\{u_1,...,u_{M}\}$ denotes the social actors, $\mathcal{E}=\{e_{ij}\}$ denotes the links between social actors, and $\mathcal{A}=\{\mathcal{A}_i\}$ denotes the attributes of social actors. 
Each edge $e_{ij}$ can be associated with a weight $s_{ij}$ denoting the strength of connection between $u_i$ and $u_j$.
Generally, our analysis can apply to any (un)directed, (un)weighted network. While in this paper, we focus on unweighted network, \textit{i.e.,} $s_{ij}$ is $1$ for all edges, our method can be easily applied to weighted network through the neighborhood sampling strategy \cite{grovernode2vec}.

\begin{figure}[t]
	\centering
	\includegraphics[scale=0.29]{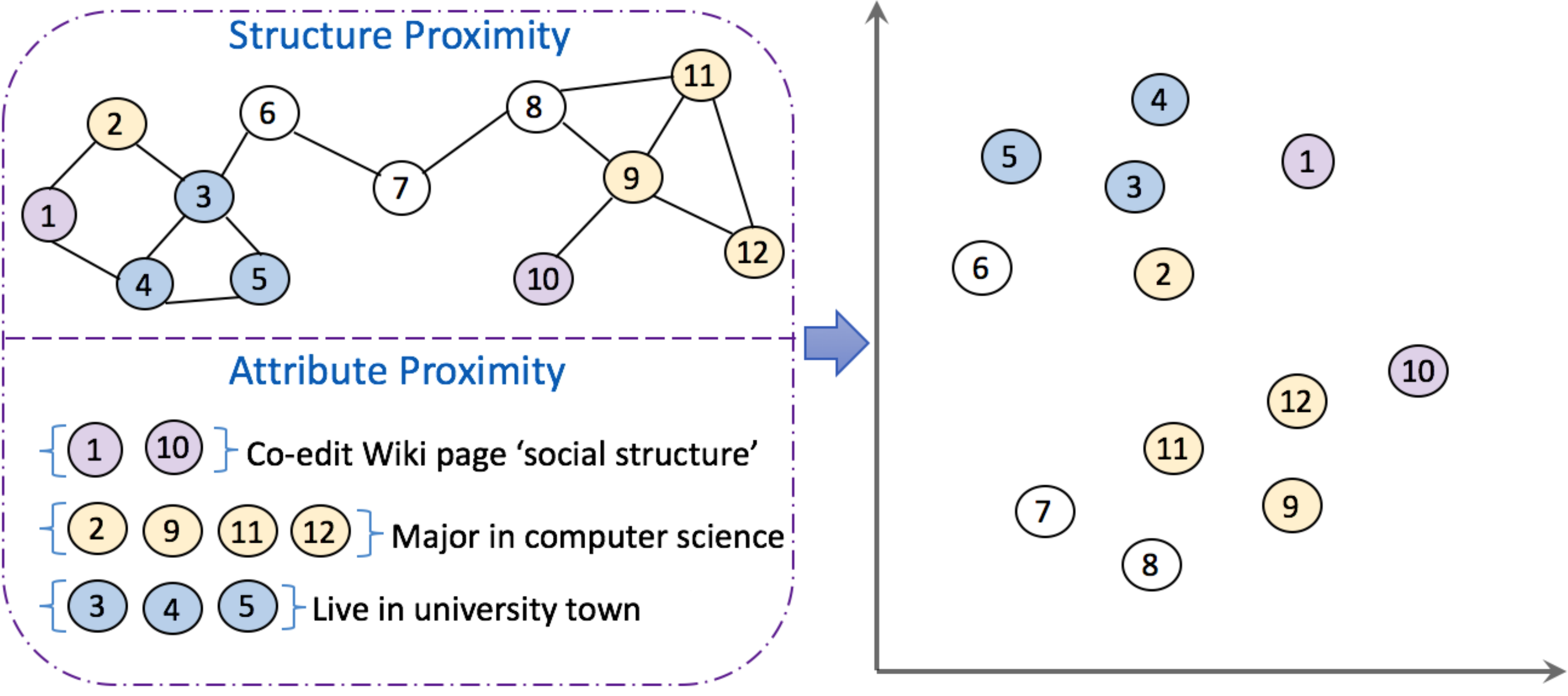}
	\caption{An illustration of social network embedding. The numbered nodes denote users, and users of the same color share the referred attribute.}
	\label{Fig:example}
\end{figure}

The aim of social network embedding is to project the social actors into a low-dimensional vector space (\textit{a.k.a.} embedding space). 
Since the network structure and attributes offer different sources of information, it is crucial to capture both of them to learn a comprehensive representation of social actors. 
To illustrate this point, we show an example in Figure~\ref{Fig:example}. 
Based on the link structure, a common assumption of network embedding methods~\cite{grovernode2vec,perozzi2014deepwalk,tang2015line} is that closely connected users should be close to each other in the embedding space. For example, $(u_1,u_2,u_3,u_4,u_5)$ should be close to each other, and similarly for $(u_8,u_9,u_{11},u_{12})$.
However, we argue that purely capturing structural information is far from enough. 
Taking the attribute homophily effect into consideration, $(u_2,u_9,u_{11},u_{12})$ should also be close to each other. 
This is because they all major in computer science; although $u_2$ is not directly linked to $u_9,u_{11}$ or $u_{12}$, we could expect that some computer science articles popular among $(u_9,u_{11},u_{12})$ might also be of interest to $u_2$. To learn more informative representations for social actors, it is essential to capture the attribute information. 

In this work, we strive to develop embedding methods that preserve both the \textit{structural proximity} and \textit{attribute proximity} of social network. In what follows, we give the definition of the two notions. 

\begin{mydef}{(Structural Proximity)} denotes the proximity of social actors that is evidenced by links. For $u_i$ and $u_j$, if there exists a link $e_{ij}$ between them, it indicates the direct proximity; on the other hand, if $u_j$ is within the context of $u_i$, it indicates the indirect proximity.
\end{mydef} 

Intuitively, the direct proximity corresponds to the first-order proximity, while the indirect proximity accounts for higher-order proximities~\cite{tang2015line}. A popular way to generate contexts is by performing random walks in the network~\cite{perozzi2014deepwalk}, \textit{i.e.,} if two nodes appear in a walking sequence, they are treated as in the same context. In our method, we apply the walking procedure proposed by \textit{node2vec}~\cite{grovernode2vec}, which controls the random walk by balancing the breadth-first sampling (BFS) and depth-first sampling (DFS). In the remaining of the paper, we use the term ``neighbors'' to denote both the first-order neighbors and the nodes in the same context for simplicity. 

\begin{mydef}{(Attribute Proximity)} denotes the proximity of social actors that is evidenced by attributes. The attribute intersection of $\mathcal{A}_i$ and $\mathcal{A}_j$  indicates the attribute proximity of $u_i$ and $u_j$.
\end{mydef} 

By enforcing the constraint of \textit{attribute proximity}, we can model the attribute homophily effect, as social actors with similar attributes will be placed close to each other in the embedding space. 

\section{Proposed Method}
We first describe how we model the \textit{structural proximity} with a deep neural network architecture. We then elaborate how to model the \textit{attribute proximity} with a similar architecture by casting attributes to a generic feature representation. Our final \textit{SNE} model integrates the models of structures and attributes by an early fusion on the input layer. Lastly, we discuss the relationships of our \textit{SNE} model to other relevant models. Some of the terms and notations are summarized in Table \ref{Table:notation}.

\subsection{Structure Modeling}
\label{ss:structure}
Since the focus of this subsection is on the modeling of network structure, we use only the identity (ID) to represent a node in the one-hot representation, in which a node $u_i$ is represented as an $M$-dimensional sparse vector where only the $i$-th element of the vector is $1$.
Based on our definition of \textit{structural proximity}, the key to structure modeling is in the estimation of pairwise proximity of nodes.
Let $f$ be the function that maps two nodes $u_i, u_j$ to their estimated proximity score. We define the conditional probability of node $u_j$ on $u_i$ using the softmax function as:
\begin{equation}
p(u_j | u_i) = \frac{exp(f(u_i, u_j))}{\sum_{j'=1}^{M} exp(f(u_i, u_{j'}))},
\label{p}
\end{equation}
which measures the likelihood that node $u_j$ is connected with $u_i$.
To account for a node's \textit{structural proximity} \textit{w.r.t.} all its neighbors, we further define the conditional probability of a node set by assuming conditional independence:
\begin{equation}
p(\mathcal{N}_i | u_i) = \prod_{j \in \mathcal{N}_i} p(u_j | u_i),
\label{eq:conditional}
\end{equation}
where $\mathcal{N}_i$ denotes the neighbor nodes of $u_i$. By maximizing this conditional probability over all nodes, we can achieve the goal of preserving the global \textit{structural proximity}. Specifically, we define the likelihood function for the global structure modeling as:
\begin{equation}
l = \prod_{i=1}^{M} p(\mathcal{N}_i | u_i) = \prod_{i=1}^{M} \prod_{j \in \mathcal{N}_i} p(u_j | u_i).
\label{likelihood}
\end{equation}

Having established the target of learning from network data, we now design an embedding model to estimate the pairwise proximity $f(u_i, u_j)$. Most previous efforts have used shallow models for relational modeling, such as matrix factorization~\cite{cao2015grarep,he2016fast} and neural networks with one hidden layer~\cite{grovernode2vec,mikolov2013efficient,perozzi2014deepwalk}.
In these formulations, the proximity of two nodes is usually modeled as the inner product of their embedding vectors.
However, It is known that simply the inner product of embedding vectors can limit the model's representation ability and incur large ranking loss \cite{he2017ncf}. To capture the complex non-linearities of real-world networks \cite{wangstructural,luo2011cauchy}, we propose to adopt a deep architecture to model the pairwise proximity of nodes:
\begin{equation}
\begin{split}
& f_{id}(u_i, u_j) \\
& =\tilde{\textbf{u}}_j\cdot \delta_n(\textbf{W}^{(n)}(\cdots\delta_1(\textbf{W}^{(1)}\textbf{u}_i +  \textbf{b}^{(1)})\cdots )+ \textbf{b}^{(n)}),
\end{split}
\label{fid}
\end{equation}
where $\textbf{u}_i$ denotes the embedding vector of node $u_i$, and $n$ denotes the number of hidden layers to transform an embedding vector to its final representation; $\textbf{W}^{(n)}$, $\textbf{b}^{(n)}$ and $\delta_n$ denote the weight matrix, bias vector and activation function of the $n$-th hidden layer, respectively.

It is worth noting that in our model design, each node has two latent vector representations, $\textbf{u}$ that encodes a node to its embedding and $\tilde{\textbf{u}}$ that embeds the node as a neighbor. To comprehensively represent a node for downstream applications, practitioners can add/concatenate the two vectors which has empirically shown to have better performance in distributed word representations~\cite{pennington2014glove,TACL570}.

\subsection{Encoding Attributes}

Many real-world social networks contain rich attribute information, which can be heterogeneous and highly diverse. To avoid manual efforts that design specific model components for specific attributes, we convert all attributes to a generic feature vector representation (see Figure~\ref{Fig:attr} as an example) to facilitate designing a general method for learning from attributes. Regardless of semantics, we can categorize attributes into two types: 

\begin{table}
	\centering
	\small
	\renewcommand{\arraystretch}{1.2}
	\caption{Terms and Notations}
	\label{Table:notation}
	\begin{tabular}{|c|l|}
		\hline
		Symbol & Definition\\
		\hline
		\hline
		M & total number of social actors in the social network  \\
		$\mathcal{N}_i$ & neighbor nodes of social actor $u_i$  \\
		$n$ & number of hidden layers\\
		$\tilde{\textbf{U}}$& the weight matrix connecting to the output layer\\
		$\textbf{h}^{(n)}_i$& embedding of $u_i$ with both structure and attributes\\
		$\tilde{\textbf{u}}_i$& the row in $\tilde{\textbf{U}}$ refers to $u_i$'s embedding as a neighbor\\
		$\textbf{u}_i$& pure structure representation of $u_i$\\
		$\textbf{u}'_i$& pure attribute representation of $u_i$ \\
		$\textbf{W}^{(k)}$, $\textbf{b}^{(k)}$ & the $k$-th hidden layer weight matrix and biases\\
		$\textbf{W}^{id}$, $\textbf{W}^{att}$ & the weight matrix for id and attributes input\\
		\hline
	\end{tabular}
\end{table}

\begin{itemize}
	\item \textbf{Discrete attributes}. A prevalent example is categorical variables, such as user demographics like gender and country. We convert a categorical attribute to a set of binary features via one-hot encoding. For example, the gender attribute has two values $\{male, female\}$, so we can express a female user as the vector $\textbf{v} = \{0,1\}$ where the second binary feature of value 1 denotes ``female''.
	\item \textbf{Continuous attributes}. Continuous attributes naturally exist on social networks, \textit{e.g.,} raw features of images and audios. Or they can be artificially generated from transformation of categorical variables. For example, in document modeling, after obtaining bag-of-words representation of a document, it is common to transform it to real-valued vector via TF-IDF to reduce noises. Another example is the historical features, such as users' purchases on items and check-ins on locations, which are always normalized to real-valued vector to reduce the impact of variable length~\cite{FM}.

	\begin{figure}[h]
		\centering
		\includegraphics[scale=0.65]{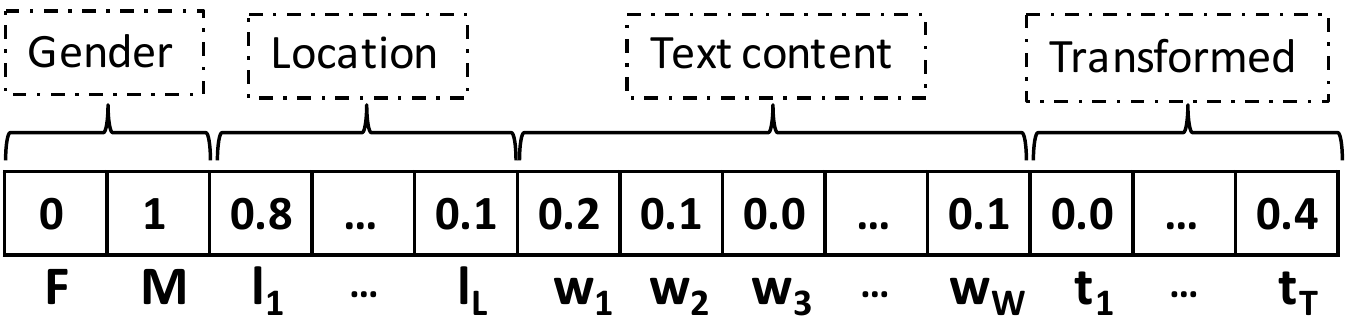}
		\caption{A simple example to show the two kinds of social network attributes information.}
		\label{Fig:attr}
	\end{figure}
	
\end{itemize}

Suppose there are $K$ feature entries in the attribute feature vector $\textbf{v}$ as shown in Figure \ref{Fig:attr}, for each feature entry, we associate it with an low-dimensional embedding vector $\textbf{e}_k$ which corresponds to the $k$-th column of the weight matrix $\textbf{W}^{att}$ as shown in Figure \ref{Fig:model}. We then aggregate the attribute representation vector $\textbf{u}'$ for each input social actor by $\textbf{u}' = \sum_{k=1}^{K}v_k\textbf{e}_k$.

Similar to structure modeling, we aim to model the \textit{attribute proximity} by adopting a deep model to approximate the complex interactions between attributes and introduce non-linearity, which can be fulfilled by Equation \ref{fid} while substituting $\textbf{u}_i$ with $\textbf{u}'_i$.

\subsection{The SNE Model}
\begin{figure}[t]
	\centering
	\includegraphics[scale=0.3]{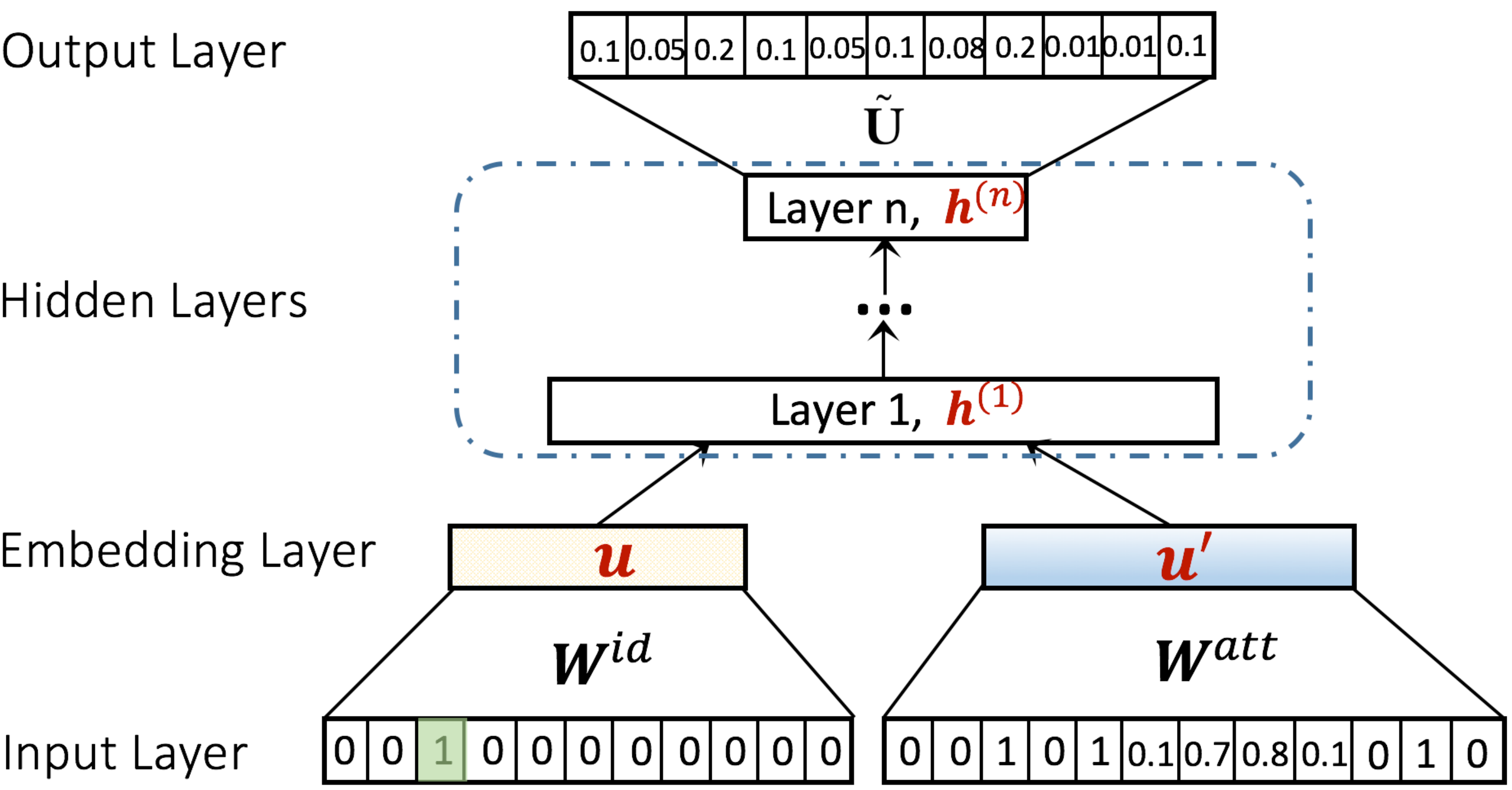}
	\caption{Social network embedding (\textit{SNE}) framework.}
	\label{Fig:model}
\end{figure}

To combine the strength of both structure and attribute modeling, an intuitive way is to concatenate the learned embeddings from each part by \textit{late fusion} as adopted by \cite{tang2015line}. However, the main drawback of late fusion is that individual models are trained separately without knowing each other and results are simply combined after training. On the contrary, \textit{early fusion} allows optimizing all parameters simultaneously. As a result, the attribute modeling can complement the learning of structure modeling, allowing teh two parts closely interact with each other. Essentially, the strategy of early fusion is more preferable in recent developments of end-to-end deep learning methods, such as Deep crossing \cite{shan2016deep} and Neural Factorization Machines \cite{he2017neu}. Therefore, we propose a generic social network embedding framework (\textit{SNE}) as shown in Figure \ref{Fig:model}, which integrates the structure and attribute modeling parts by an early fusion on the input layer. In what follows, we elaborate the design of \textit{SNE} layer by layer.

\textbf{Embedding Layer}. The embedding layer consists of two fully connected components. One component projects the one-hot user ID vector to a dense vector $\textbf{u}$ which captures structure information. The other component encodes the generic feature vector and generates a compact vector $\textbf{u}'$ which aggregates attributes information.

\textbf{Hidden Layers}. Above the embedding layer, $\textbf{u}$ and $\textbf{u}'$ are fed into a multi-layer perceptron. The hidden representations for each layer are denoted as $\textbf{h}^{(0)}, \textbf{h}^{(1)}, \cdots ,\textbf{h}^{(n)}$, which are defined as follows:
\begin{equation}
\begin{split}
& \textbf{h}^{(0)} = \begin{bmatrix}
\textbf{u}\\
\lambda\textbf{u}'
\end{bmatrix},\\
& \textbf{h}^{(k)} = \delta_k ( \textbf{W}^{(k)}\textbf{h}^{(k-1)} + \textbf{b}^{(k)}),~k = 1, 2,\cdots, n,
\end{split}
\label{concate}
\end{equation}
where $\lambda \in \mathbb{R}$ adjusts the importance of attributes, $\delta_k$ denotes the activation function, $n$ is the number of hidden layers. From the last hidden layer, we obtain an abstractive representation $\textbf{h}^{(n)}_i$ of the input social actor $u_i$.

Stacking multiple non-linear layers has been shown to help learning better representations of data \cite{he2015deep}. Regarding the architecture design, a common strategy is to use a tower structure, where each successive layer has a smaller number of neurons. The premise is that by using a small number of hidden units for higher layers, they can learn more abstractive features of data \cite{he2015deep}. Therefore, as depicted in Figure \ref{Fig:model}, we implement the \textit{hidden layers} component following the tower structure with halved layer size for each successive higher layer. Such a design has also been shown to be effective by recent work on recommendation task \cite{he2017ncf}. Moreover, $\textbf{u}$ and $\textbf{u}'$ are concatenated with weight adjustments $\lambda$ before fed into the fully connected layers, which can help to learn high-order interactions between also has been shown to help learning higher-order interactions between $\textbf{u}$ and $\textbf{u}'$\cite{shan2016deep,he2017ncf}.

\textbf{Output Layer}. At last, the output vector of the last hidden layer $\textbf{h}^{(n)}_i$ is transformed into a probability vector $\textbf{o}$, which contains the predictive link probability of $u_i$ to all the nodes in $\mathcal{U}$:
\begin{equation}
\textbf{o} = [ p(u_1|u_i), p(u_2|u_i),\cdots, p(u_{M}|u_i) ].
\end{equation}
Denoting the abstractive representation of a neighbor $u_j$ as $\tilde{\textbf{u}}_j$ which corresponds to a row in the weight matrix $\tilde{\textbf{U}}$ between the last hidden layer and the output layer, the proximity score between $u_i$ and $u_j$ can be defined as below:
\begin{equation}
f(u_i,u_j) = \tilde{\textbf{u}}_j \cdot \textbf{h}^{(n)}_i,
\label{f}
\end{equation}
which can be fed into Equation \ref{p} for further obtaining the predictive link probability $p(u_j|u_i)$ in vector $\textbf{o}$:
\begin{equation}
p(u_j | u_i) = \frac{exp(\tilde{\textbf{u}}_j \cdot \textbf{h}^{(n)}_i)}{\sum_{j'=1}^{M} exp(\tilde{\textbf{u}}_j' \cdot \textbf{h}^{(n)}_i)},
\end{equation}
where all the parameters $\Theta=\{\Theta_h,\textbf{W}^{id},\textbf{W}^{att}, \tilde{\textbf{U}}\}$ and $\Theta_h$ denotes the weight matrices and biases in the \textit{hidden layers} component.

\subsubsection{Optimization}
To estimate the model parameters of the whole \textit{SNE} framework, we need to specify an objective function to optimize. As detailed in Equation \ref{likelihood}, we aim to maximize the conditional link probability over all nodes. In this way, the whole \textit{SNE} framework is jointly trained to maximize the likelihood with respect to all the parameters $\Theta$,
\begin{align}
\Theta^\star &= \argmax_{\Theta} \prod_{i=1}^{M}\prod_{j \in \mathcal{N}_i} p(u_j | u_i) \notag \\
&=\argmax_{\Theta} \sum_{u_i \in {M}} \sum_{u_j \in \mathcal{N}_i }log~p(u_j|u_i) \label{Equ:logp}\\
&= \argmax_{\Theta} \sum_{u_i \in {M}} \sum_{u_j \in \mathcal{N}_i }log~\frac{exp(\tilde{\textbf{u}}_j\cdot \textbf{h}^{(n)}_i)}{\sum_{j'\in {M}} exp(\tilde{\textbf{u}}_{j'}\cdot \textbf{h}^{(n)}_i)}.
\label{Equ:obj}
\end{align}

Maximizing the softmax scheme in Equation \ref{Equ:obj} actually has two effects: to enhance the similarity between any $u_i$ and these $u \in \mathcal{N}_i$ as well as to weaken that between any $u_i$ and these $u \not \in \mathcal{N}_i$. However, this causes two major problems. The first one lies in the fact that if two social actors are not linked together, it does not necessarily mean they are dissimilar. For example, many users in social websites are not linked, not because they are dissimilar. Most of the times, it is simply because they never had the chance to know each other. Thus forcing dissimilarity between $u_i$ and all the other actors not inside $\mathcal{N}_i$ will be inappropriate. The second problem arises from the calculation of the normalization constant in Equation \ref{Equ:obj}. In order to calculate a single probability, we need to go through all the actors in the whole network, which is computationally inefficient. In order to avoid these problems, we apply negative sampling procedure \cite{mikolov2013efficient,wang2017your} where only a very small subset of users are sampled from the whole social network.

The main idea is to do approximation in the gradient calculation procedure. When we consider the gradient of the log-probability in Equation \ref{Equ:logp}, the gradient is actually composed of a positive and a negative part as follows,
\begin{equation*}
\vspace{-0.1cm}
\nabla~log~p(u_j|u_i) = \nabla~f(u_i, u_j)-\sum_{j'\in{M} }p(u_{j'}|u_i)\nabla~f(u_i, u_{j'}),
\label{Equation:grad}
\vspace{-0.1cm}
\end{equation*}
where $f(u_i,u_j) = \tilde{\textbf{u}}_j\cdot \textbf{h}^{(n)}_i$ as defined in Equation \ref{f}. Note that given the actor $u_i$, the negative part of the gradient is in essence the expected gradient of $\nabla f(u_i, u_{j'})$, denoting as $\mathbb{E}[\nabla f(u_i, u_{j'})]$. The key idea for sampling a subset of social actors is to approximate this expectation, resulting in much lower computational complexity as well as avoiding too strong constraint on those not linked actors.

To optimize the aforementioned framework, we apply the Adaptive Moment Estimation (Adam) \cite{kingma2014adam}, which adapts the learning rate for each parameter by performing smaller updates for the frequent parameters and larger updates for the infrequent parameters. The Adam method combines the advantages of two popular optimization methods: the ability of AdaGrad \cite{duchi2011adaptive} to deal with sparse gradients, and the ability of RMSProp \cite{tieleman2012lecture} to deal with non-stationary objectives. To address internal covariate shift \cite{IoffeS15} which slows down the training by requiring careful settings of learning rate and parameter initialization, we adopt batch normalization \cite{IoffeS15} in our multi-layer \textit{SNE} framework. In the embedding layer and each hidden layer, we also add dropout component to alleviate overfitting. After proper optimization, we obtain abstractive representation $\textbf{h}^{(n)}$ and $\tilde{\textbf{u}}$ for each social actor, similar to \cite{pennington2014glove,TACL570}, we use $\textbf{h}^{(n)}+\tilde{\textbf{u}}$ as the final representation for each social actor, which returns us better performance results.

\subsection{Connections to Other Models}	
In this subsection, we discuss the connection of the proposed \textit{SNE} framework to other related models. We show that \textit{SNE} subsumes the state-of-the-art network embedding method \textit{node2vec} \cite{grovernode2vec}  and the linear latent factor model \textit{SVD++}~\cite{koren2008factorization}.
Specially, the two models can be seen as a special case of shallow \textit{SNE}. 
To facilitate further discussion, we first give the prediction model of the one-hidden-layer \textit{SNE} as:

\begin{equation}
f(u_i, u_j)
=\tilde{\textbf{u}}_j\cdot \delta_1(\textbf{W}^{(1)}
\begin{bmatrix}
\textbf{u}_i\\
\lambda\textbf{u}'_i
\end{bmatrix}\\
+  \textbf{b}^{(1)}).
\label{target}
\end{equation}

\subsubsection{\textit{SNE} vs. \textit{node2vec}} The \textit{node2vec} applies a shallow neural network model to learning node embeddings. Under the context of \textit{SNE}, the essence of \textit{node2vec} can be seen as estimating the proximity of two nodes as:
\begin{equation*}
f_{node2vec}(u_i, u_j)
=\tilde{\textbf{u}}_j\cdot\textbf{u}_i.
\end{equation*}
By setting $\lambda$ to 0.0 (\textit{i.e.}, no attribute modeling), $\delta_1$ to an identity function (\textit{i.e.}, no nonlinear transformation),  $\textbf{W}^{(1)}$ to an identity matrix and $\textbf{b}^{(1)}$ to a zero vector (\textit{i.e.}, no trainable hidden neurons), we can exactly recover the \textit{node2vec} model from Equation \ref{target}.

\subsubsection{\textit{SNE} vs. \textit{SVD++}} The \textit{SVD++} is one of the most effective latent factor models for collaborative filtering~\cite{koren2008factorization}, originally proposed to model the ratings of users to items. 
Given a user $u$ and an item $i$, the prediction model of \textit{SVD++} is defined as:
\begin{equation*}
f_{SVD++}(u, i) = \textbf{q}_i \cdot \left(\textbf{p}_u + \sum_{k\in \mathcal{R}_u}\textbf{y}_k\right),
\end{equation*}
where $\textbf{p}_u$ ($\textbf{q}_i$) denotes the embedding vector for user $u$ (item $i$); $\mathcal{R}_u$ denotes the set of rated items for $u$, and $\textbf{y}_k$ denotes another embedding vector for item $k$ for modeling the item--item similarity. 
By treating the item as a ``neighbor'' of the user for estimating the proximity, we reformulate the model using the symbols of our \textit{SNE}: 
\begin{equation*}
f_{SVD++}(u_i, u_j) = \tilde{\textbf{u}}_j\cdot \left(\textbf{u}_i + \textbf{u}'_i\right),
\end{equation*}
where $\textbf{u}'_i$ denotes the sum of item embedding vectors of $\mathcal{R}_u$, which corresponds to the aggregated attribute representation of $u_i$ in \textit{SNE}. 

To see how \textit{SNE} subsumes the model, we first set $\delta_1$ to an identity function, $\lambda$ to $1.0$, and $\textbf{b}^{(1)}$ to a zero vector, reducing Equation \ref{target} to:
\begin{equation*}
f(u_i,u_j)
= \tilde{\textbf{u}}_j \cdot \textbf{W}^{(1)}\begin{bmatrix}
\textbf{u}_i\\
\textbf{u}'_i
\end{bmatrix}.
\end{equation*}

By further setting $\textbf{W}^{(1)}$ to a concatenation of two identity matrices (\textit{i.e.} $\textbf{W}^{(1)} = [\textbf{I, I}]$), we can recover the \textit{SVD++} model:
\begin{equation*}
f(u_i,u_j) = \tilde{\textbf{u}}_j\cdot \left(\textbf{u}_i + \textbf{u}'_i\right).
\end{equation*} 

\noindent Through the connection between \textit{SNE} and a family of shallow models, we can see the rationality behind our design of \textit{SNE}. Particularly, \textit{SNE} deepens the shallow models so as to capture the underlying interactions between the network structure and attributes. When modeling real-world data that may have complex and non-linear inherent structure~\cite{wangstructural,luo2011cauchy}, our \textit{SNE} is more expressive and can better fit on the real-world data.

\section{Experiments}
In this section, we conduct experiments on four publicly accessible social network datasets to answer the following research questions.
\begin{description}
	\item [RQ1] Can \textit{SNE} learn better node representations as compared to  state-of-the-art network embedding methods?
	\item [RQ2] What are the key reasons that lead to better representations learned by \textit{SNE}?
	\item [RQ3] Are deeper layers of hidden units helpful for learning better social network embeddings?
\end{description}
In what follows, we first describe the experimental settings. We then answer the above three research questions one by one.

\subsection{Experimental Setup}
\subsubsection{Datasets}
\label{ss:datasets}
We conduct the experiments on four public datasets, which are representative of two types of social networks --- social friendship networks and academic citation networks \cite{wasserman1994social}. The statistics of the four datasets are summarized in Table~\ref{Table:datainfo}.

\textbf{FRIENDSHIP Networks}. We use two Facebook networks constructed by \cite{traud2012social}, which contain students from two American universities: University of Oklahoma (OKLAHOMA) and University of North Carolina at Chapel Hill (UNC), respectively.
Besides user ID, there are seven anonymized attributes: status, gender, major, second major, dorm/house, high school, class year. Note that not all students have the seven attributes available. For example, for the UNC dataset, only $4,018$ of the $18,163$ users contain all attributes (as plotted in Figure~\ref{Fig:block}).

\textbf{CITATION Networks}. For citation networks, we use the DBLP and CITESEER\footnote{http://citeseerx.ist.psu.edu/} data used in \cite{shirui2016}.
Each node denotes a paper. The attributes are the title contents for each paper after removing stop words and the stemming process. The DBLP dataset consists of bibliography data in computer science from \cite{tang2008}\footnote{http://arnetminer.org/citation (V4 version is used)}.
A list of conferences from four research areas are selected. The CITESEER dataset consists of scientific publications from ten distinct research areas. These research areas are treated as class labels in the node classification task.

\begin{table}[!htp]
	\centering
	\small
	\renewcommand{\arraystretch}{1.35}
	\caption{Statistics of the datasets}
	\label{Table:datainfo}
	\begin{tabular}{|c|c|c|}
		\hline
		Dataset& \#($\mathcal{U}$) & \#($\mathcal{E}$)\\
		\hline
		OKLAHOMA~\cite{traud2012social} &~~~~17,425~~~~&~~~~892,528~~~~\\
		\hline
		UNC~\cite{traud2012social} &~~~~18,163~~~~&~~~~766,800~~~~\\
		\hline
		DBLP~\cite{shirui2016}&~~~~60,744~~~~&~~~52,890~~~~~\\
		\hline
		CITESEER~\cite{shirui2016} &~~~~29,751~~~~&~~~~77,218
		~~~~ \\ 
		\hline
	\end{tabular}
\end{table}
\subsubsection{Evaluation Protocols}
\label{evalproto}
We adopt two tasks --- link prediction and node classification --- 
which have been widely used in literature to evaluate network embeddings~\cite{perozzi2014deepwalk,grovernode2vec}. 
While the link prediction task assesses the ability of node representations in reconstructing network structure \cite{wangstructural}, node classification evaluates whether the representations contain sufficient information trainable for downstream applications.

\textbf{Link prediction}. We follow the widely adopted way in \cite{grovernode2vec,wangstructural}: we randomly hold out $10\%$ links as the test set, $10\%$ as the validation set for tuning hyper-parameters, and train \textit{SNE} on the remaining $80\%$ links. 
Since the test/validation set contains only positive instances, we randomly sample the same number of non-existing links as negative instances~\cite{grovernode2vec}, and rank both positive and negative instances according to the prediction function. 
To judge the ranking quality, 
we employ the \textit{area under the ROC curve} (AUROC)~\cite{zou2007receiver}, which is widely used in IR community to evaluate a ranking list. 
It is a summary measure that essentially averages accuracy across the spectrum of test values.
A higher value indicates a better performance, and an ideal model that ranks all positive instances higher than negative instances has an AUROC value of 1. 

\textbf{Node classification}. 
We first train models on the training sets (with links and all attributes but no class labels) to obtain node representations; the hyper-parameters for each model are chosen based on the performance of link prediction. 
We then feed node representations into the LIBLINEAR package~\cite{fan2008liblinear}, which is widely adopted in \cite{perozzi2014deepwalk,wangstructural}, to train a classifier.
To evaluate the classifier, 
we randomly sample a portion of labeled nodes ($\rho \in \{10\%, 30\%, 50\%\}$) as training, using the remaining labeled nodes as test. 
We repeat this process 10 times, and report the mean of the \textit{Macro-F1} and \textit{Micro-F1} scores. 
Note that since only the DBLP and CITESEER datasets contain class labels for nodes, the node classification task is performed on the two datasets only. 

\subsubsection{Comparison Methods}
\label{methods}
We compare \textit{SNE} with several state-of-the-art network embedding methods. \\

\textbf{\textit{- node2vec}}~\cite{grovernode2vec}: It applies the Skip-Gram model~\cite{mikolov2013efficient} on the node sequences generated by biased random walk. There are two key hyper-parameters $p$ and $q$ that control the random walk, which we tuned them the same way as the original paper. Note that when $p$ and $q$ are set to 1, \textit{node2vec} degrades to
DeepWalk \cite{perozzi2014deepwalk}.   

\textbf{\textit{- LINE}}~\cite{tang2015line}: It learns two embedding vectors for each node by preserving the first-order and second-order proximity of the network, respectively. Then the embedding vectors are concatenated as the final representation for a node. 
We followed the hyper-parameter settings of \cite{tang2015line} and the number of training samples $S$ (millions) 
is adapted to our data size.

\textbf{\textit{- TriDNR}}~\cite{shirui2016}: 
It learns node representations by coupling multiple neural network models
to jointly exploit the network structure, node--content correlation, and label--content correspondence. 
This is a state-of-the-art network embedding method that also uses attribute information. 
We searched the text weight ($tw$) hyper-parameter among $[0.0,0.2,...,1.0]$. 

For all baselines, we used the implementation released by the original authors.
Note that although \textit{node2vec} and \textit{LINE} are state-of-the-art methods for embedding networks, they are designed to use only the structure information. For a fair comparison with \textit{SNE} that additionally exploits attributes, we further extend them to include attributes by concatenating the learned node representation with the attribute feature vector. We dub the variants \textit{node2vec+} and \textit{LINE+}. 
Moreover, we are aware of a recent network embedding work \cite{huang2017label} also considering attribute information. However, due to the unavailability of their codes, we do not further compare with it.

\subsubsection{Parameter Settings}
\label{param}
\begin{table}[t]
	\footnotesize
	\centering
	\renewcommand{\arraystretch}{1.35}
	\caption{The optimal hyper-parameter settings.}
	\label{table:param}
	\begin{tabular}{|c|c|c|c|c|c|}
		\hline
		\multicolumn{2}{|c|}{} & OKLAHOMA  & UNC & DBLP & CITESEER \\
		\hline
		\multicolumn{1}{|c|}{\multirow{3}{*}{\textit{SNE}}}
		&$bs$&128&256&128&64\\
		\cline{2-6}
		&$lr$&0.0001&0.0001&0.001&0.001\\
		\cline{2-6}
		&$\lambda$&0.8&0.8&1.0&1.0\\
		\hline
		\multicolumn{1}{|c|}{\multirow{2}{*}{\textit{node2vec}}}
		&$p$&2.0&2.0&1.0&2.0\\
		\cline{2-6}
		&$q$&0.25&1.0&0.25&0.125\\
		\hline
		\multicolumn{1}{|c|}{\multirow{1}{*}{\textit{LINE}}}
		&$S$&100&100&10&10\\
		\hline
		\multicolumn{1}{|c|}{\multirow{1}{*}{\textit{TriDNR}}}
		&$tw$&0.6&0.6&0.8&0.8\\
		\hline
	\end{tabular}
\end{table}

\begin{figure*}[!htp]
	\centering
	\subfigure[][OKLAHOMA]{\includegraphics[scale = 0.425]{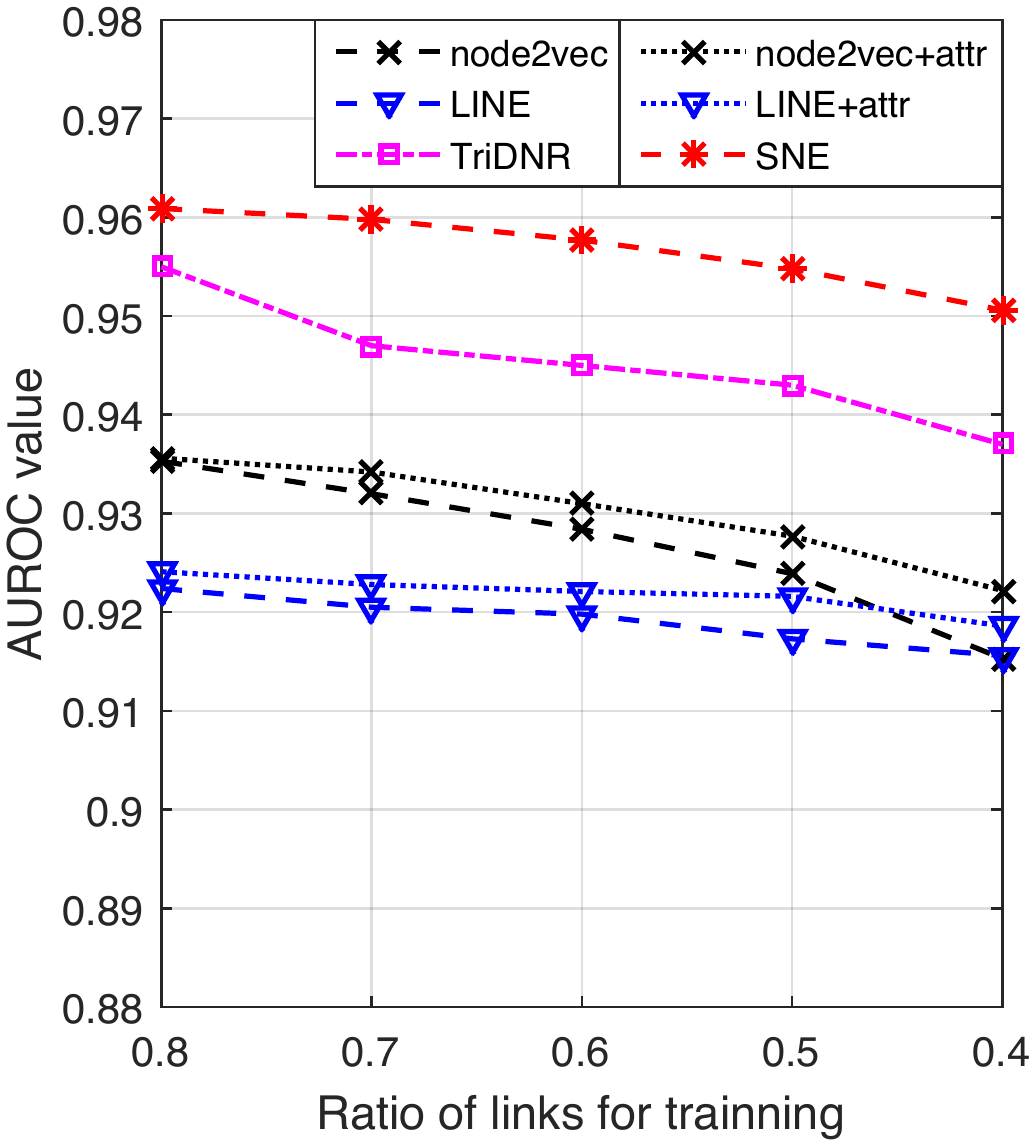}}
	\subfigure[][UNC]{\includegraphics[scale = 0.425]{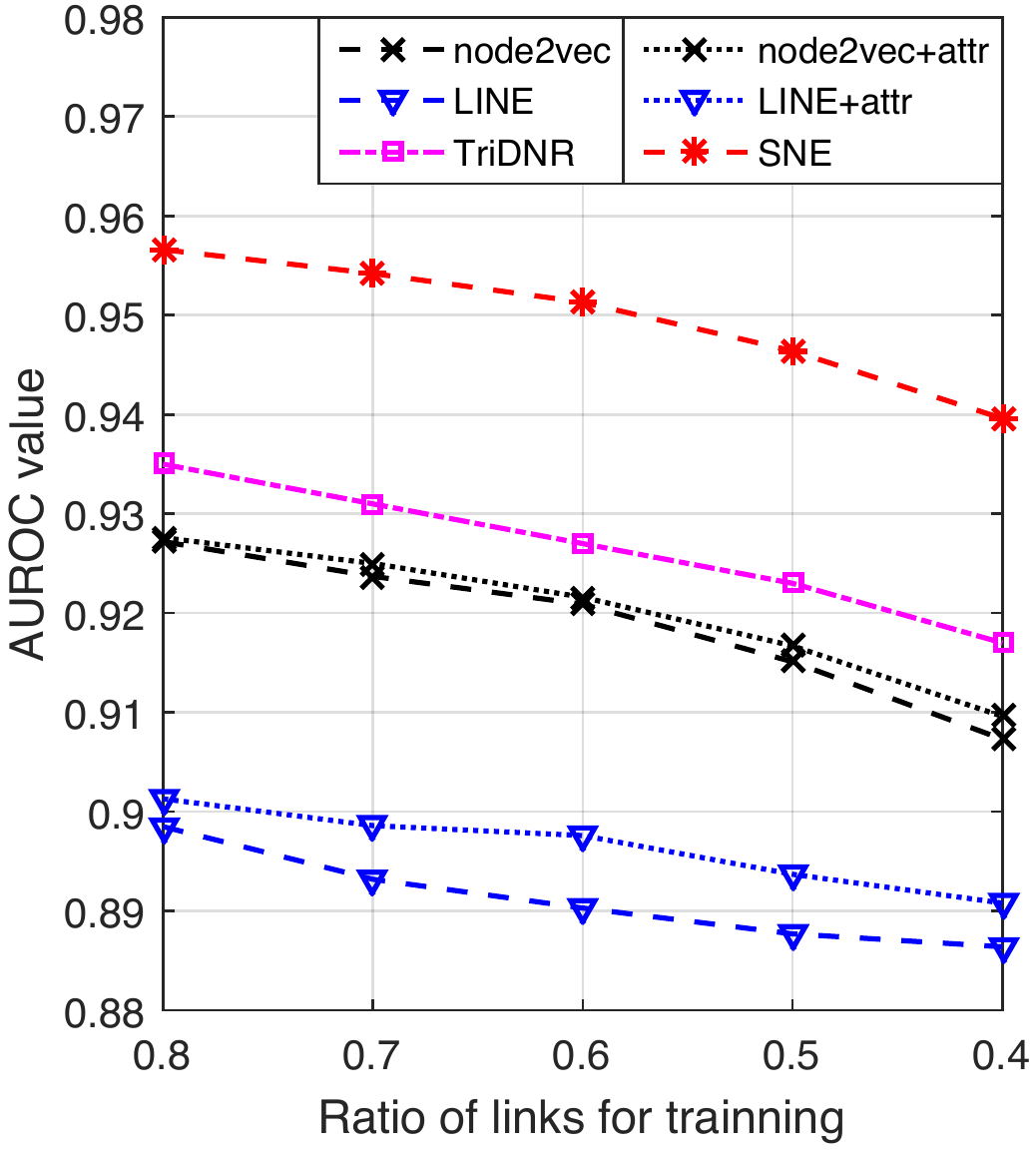}}
	\subfigure[][DBLP]{\includegraphics[scale = 0.425]{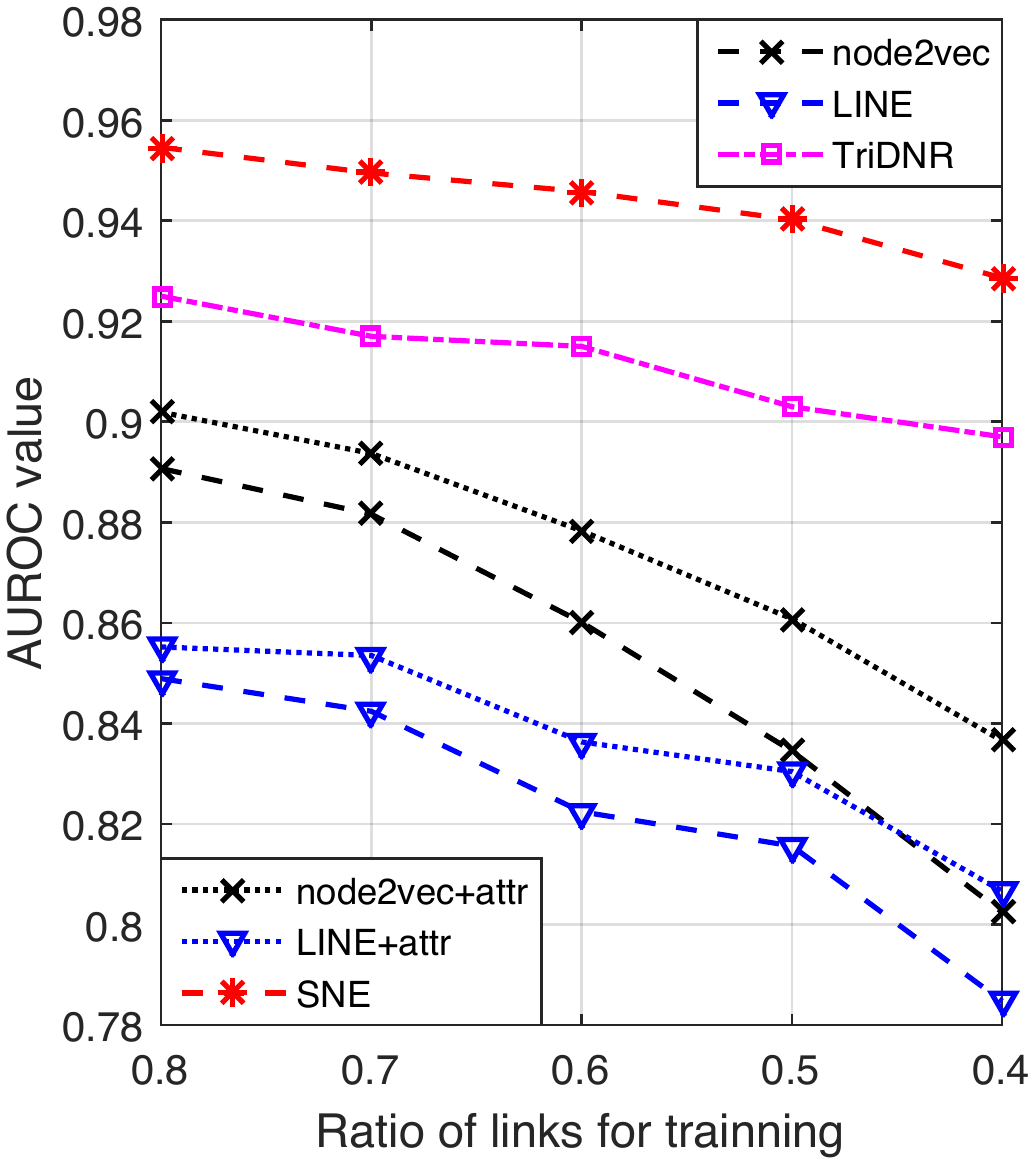}}
	\subfigure[][CITESEER]{\includegraphics[scale = 0.425]{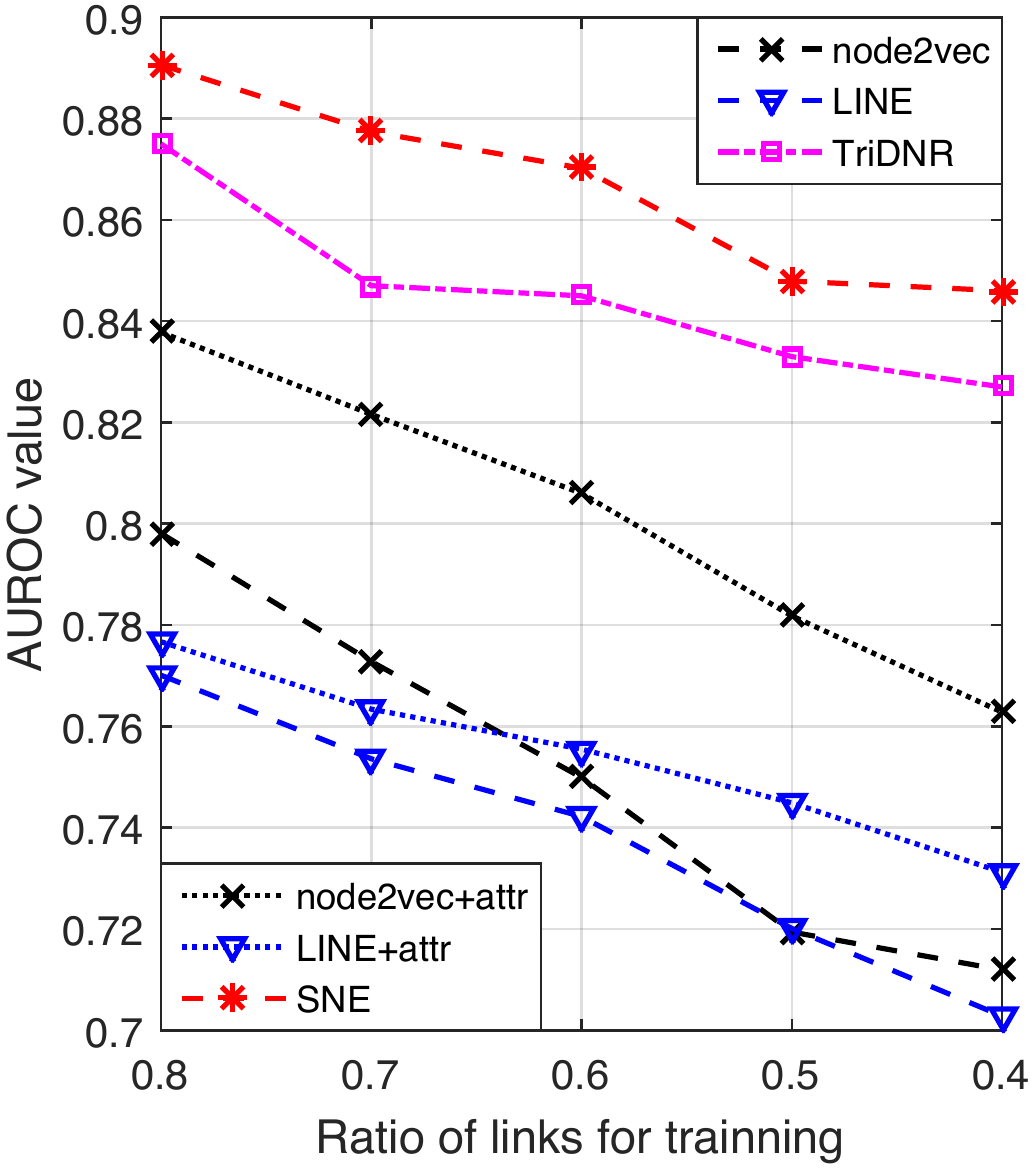}}
	\caption{Performance of link prediction on social networks \textit{w.r.t.} different network sparsity (RQ1).}
	\label{fig:linkprediction}
\end{figure*}
Our implementation of \textit{SNE} is based on TensorFlow\footnote{https://www.tensorflow.org/}, which will be made available upon acceptance.
Regarding the choice of activation function of hidden layers, we have tried rectified linear unit~(\textit{ReLU}), soft sign (\textit{softsign}) and hyperbolic tangent function (\textit{tanh}), finding softsign leads to the best performance in general. As such, we use softsign for all experiments.  
We randomly initialize model parameters with a Gaussian distribution (with a mean of 0.0 and standard deviation of 0.01), optimizing
the model with mini-batch Adam \cite{kingma2014adam}. 
We test the batch size ($bs$) of $[8,16,32,64,128, 256]$ and the learning rate ($lr$) of $[0.1,0.01,0.001,0.0001]$.
The search space of the concatenation hyper-parameter $\lambda$ is the same as $tw$ of \textit{TriDNR}, where a value of $\lambda=0.0$ degrades to a model that considers only the structure (\textit{c.f.,} Section~\ref{ss:structure}).
The concatenation parameter $\lambda$ is searched in same space as $tw$. More detailed impact of $\lambda$ is studied in Section~\ref{lambda}.
The embedding dimension $d$ is set to 128 for all methods in line with \textit{node2vec} and \textit{LINE}. The hyper-parameter $p$ and $q$ for controlling the walking procedure are set to be the same with that of \textit{node2vec}.
Without special mention, we use two hidden layers, \textit{i.e.,} $n=2$. 
Table \ref{table:param} summarizes the optimal hyper-parameters of each method tuned on validation sets.

\subsection{Quantitative Analysis (RQ1)}

\subsubsection{Link Prediction}
Figure \ref{fig:linkprediction} shows the AUROC scores of \textit{SNE} and baseline methods on the four datasets. To explore the robustness of embedding methods \textit{w.r.t.} the network sparsity, we vary the ratio of training links and investigate the performance change. The key observations are as follows:

1) Our proposed \textit{SNE} achieves the best performance among all methods. 
Notably, compared to the pure structure-based methods \textit{node2vec} and \textit{LINE}, our \textit{SNE} performs significantly better with only half links. 
This demonstrates the usefulness of attributes in predicting missing links, as well as the rationality of \textit{SNE} in leveraging attributes for learning better node representation. 
Moreover, we observe more dramatic performance drop of \textit{node2vec} and \textit{LINE} on DBLP and CITESEER, as compared to that of OKLAHOMA and UNC. 
The reason is that the DBLP and CITESEER datasets contain less link information (as shown in Table~\ref{Table:datainfo}); as such, the link sparsity problem becomes more severe when the ratio of training links decreases. 
On the contrary, our \textit{SNE} exhibits more stability when we use fewer links for training, which is credible to its effective modeling of attributes.

2) Focusing on methods that account for attributes, we find how to incorporate attributes plays a pivotal role for the performance. 
First, \textit{node2vec+} (\textit{LINE+}) slightly improves over \textit{node2vec} (\textit{LINE}), which reflects the value of attributes.
Nevertheless, the rather modest improvements indicate that simply concatenating attributes with the embedding vector is insufficient to fully leverage the rich signal in attributes. 
This reveals the necessity of designing a more principled approach to incorporate attributes into the network embedding process. 
Second, we can see that \textit{SNE} consistently outperforms
\textit{TriDNR} --- the most competitive baseline that also incorporates attributes into the network embedding process.
Although \textit{TriDNR} is a joint model, it separately trains the structured-based \textit{DeepWalk} and attributed-fused \textit{Doc2Vec} during the optimization process, which can be sub-optimal to leverage attributes. 
In contrast, our \textit{SNE} seamlessly incorporates attributes by an early fusion on the input layer, which allows the following hidden layers to capture complex structure--attribute interactions and learn more informative node representations.

3). Comparing the two structure-based methods, we observe that \textit{node2vec} generally outperforms \textit{LINE} across all the four datasets. This result is in consistent with Grover and Leskovec~\cite{grovernode2vec}'s finding. 
One plausible reason for \textit{node2vec}'s superior performance might be that by performing random walks on the social network, higher-order proximity information can be captured.
In contrast, \textit{LINE} only models the first- and second-order proximities, which fails in capturing sufficient information for link prediction. 
To justify this, we have further explored an additional baseline that directly utilizes the second-order proximity by ranking nodes according to their common neighbors. As expected, the performance is weak for all datasets~(lower than the bottom line of each subfigure), which again demonstrates the need for learning higher-order proximities via network embedding.
Since our \textit{SNE} shares the same walking procedure as \textit{node2vec}, it is also capable of learning from higher-order proximities, which are further complemented by the attribute information.

\subsubsection{Node Classification}
Table~\ref{table:nodeclass} shows the \textit{macro-F1} and \textit{micro-F1} scores obtained by each method on the classification task.
Upon getting the node representations, we train the LIBLINEAR classifier with different ratios of labeled data ($\rho \in \{10\%, 30\%, 50\%\}$). 
The performance trends are generally consistent with that of the link prediction task. 

\begin{table*}[!htp]
	\centering
	\footnotesize
	\renewcommand{\arraystretch}{1.45}
	\caption{Averaged \textit{Macro-F1}, \textit{Micro-F1} scores for node classification task. $^\star$ denotes the statistical significance for $p < 0.05$. (RQ1)}
	\label{table:nodeclass}
	\begin{tabular}{|c|c|c|c|c|c|c|c||c|c|c|c|c|c|}
		\hline
		\multicolumn{2}{|c|}{Dataset} & \multicolumn{6}{|c||}{CITESEER} & \multicolumn{6}{|c|}{DBLP} \\
		\hline
		\multicolumn{2}{|c|}{Method}&\textit{LINE}&\textit{node2vec}& \textit{LINE+}&\textit{node2vec+}&\textit{TriDNR}&\textit{SNE}&\textit{LINE}&\textit{node2vec}&\textit{LINE+}&\textit{node2vec+}&\textit{TriDNR}&\textit{SNE}\\
		\hline
		\multicolumn{1}{|c|}{\multirow{3}{*}{\rotatebox{90}{\textit{macro}}}}
		&\textit{10\%}&0.548&0.606&0.597&0.613&0.618&\textbf{0.653}$^\star$&0.565&0.617&0.619&0.631&0.665&\textbf{0.699}$^\star$\\
		\cline{2-14}
		&\textit{30\%}&0.580&0.625&0.631&0.630&0.692&\textbf{0.715}$^\star$&0.586&0.632&0.636&0.642&0.702&\textbf{0.725}$^\star$\\
		\cline{2-14}
		&\textit{50\%}&0.619&0.667&0.670&0.682&0.736&\textbf{0.752}$^\star$&0.628&0.677&0.692&0.695&0.715&\textbf{0.761}$^\star$\\
		\hline
		\hline
		\multicolumn{1}{|c|}{\multirow{3}{*}{\rotatebox{90}{\textit{micro}}}}
		&\textit{10\%}&0.573&0.623&0.607&0.628&0.644&\textbf{0.675}$^\star$&0.587&0.647&0.661&0.686&0.750&\textbf{0.763}$^\star$\\
		\cline{2-14}
		&\textit{30\%}&0.614&0.653&0.667&0.695&0.714&\textbf{0.732}$^\star$&0.632&0.665&0.678&0.749&0.778&\textbf{0.786}$^\star$\\
		\cline{2-14}
		&\textit{50\%}&0.661&0.695&0.691&0.717&0.756&\textbf{0.767}$^\star$&0.678&0.733&0.732&0.753&0.785&\textbf{0.804}$^\star$\\
		\hline
	\end{tabular}
\end{table*}

First and foremost, \textit{SNE} achieves the best performance among all the methods for all settings, and the 
one-sample paired t-test verifies that all improvements are statistically significant for $p<0.05$.
The performance of \textit{SNE} is followed by that of \textit{TriDNR}, and then followed by that of the attribute-based methods  \textit{node2vec+} and \textit{LINE+}; \textit{node2vec} and \textit{LINE} which use only the network structure perform the worst. 
This further justifies the usefulness of attributes on social networks, and such that properly modeling them can lead to better representation learning and benefit downstream applications. Among the four attribute-based methods, \textit{SNE} and \textit{TriDNR} demonstrate superior performance over \textit{node2vec+} and \textit{LINE+}, which points to the positive effects of incorporating attributes into the network embedding process. 

It is worth pointing out that the ground-truth labels of the node classification task are not involved in the network embedding process. 
Despite this, \textit{SNE} can learn effective representations that support the task well. 
This is attributed to \textit{SNE}'s modeling of network structure and attributes in a sound way, which leads to comprehensive and informative representations for nodes. 

\subsubsection{Impact of $\lambda$}
\label{lambda}
\begin{figure}[!htp]
	\centering
	\subfigure[][Link prediction]{\includegraphics[scale = 0.7]{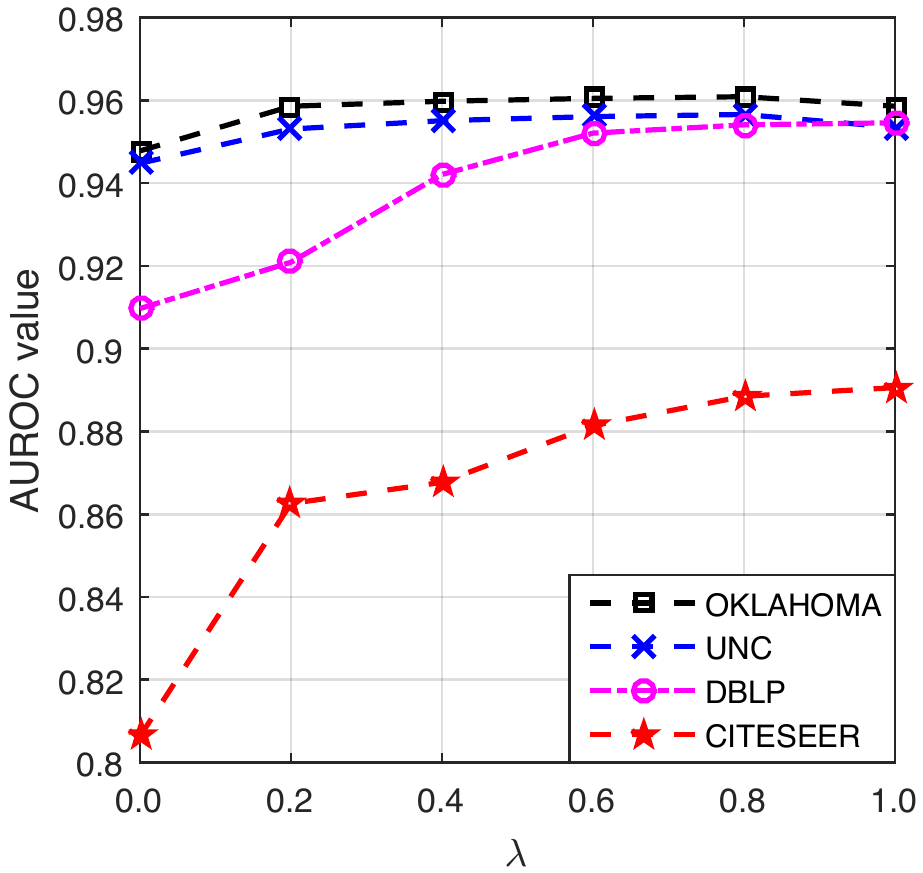}}
	\subfigure[][Node classification]{\includegraphics[scale = 0.35]{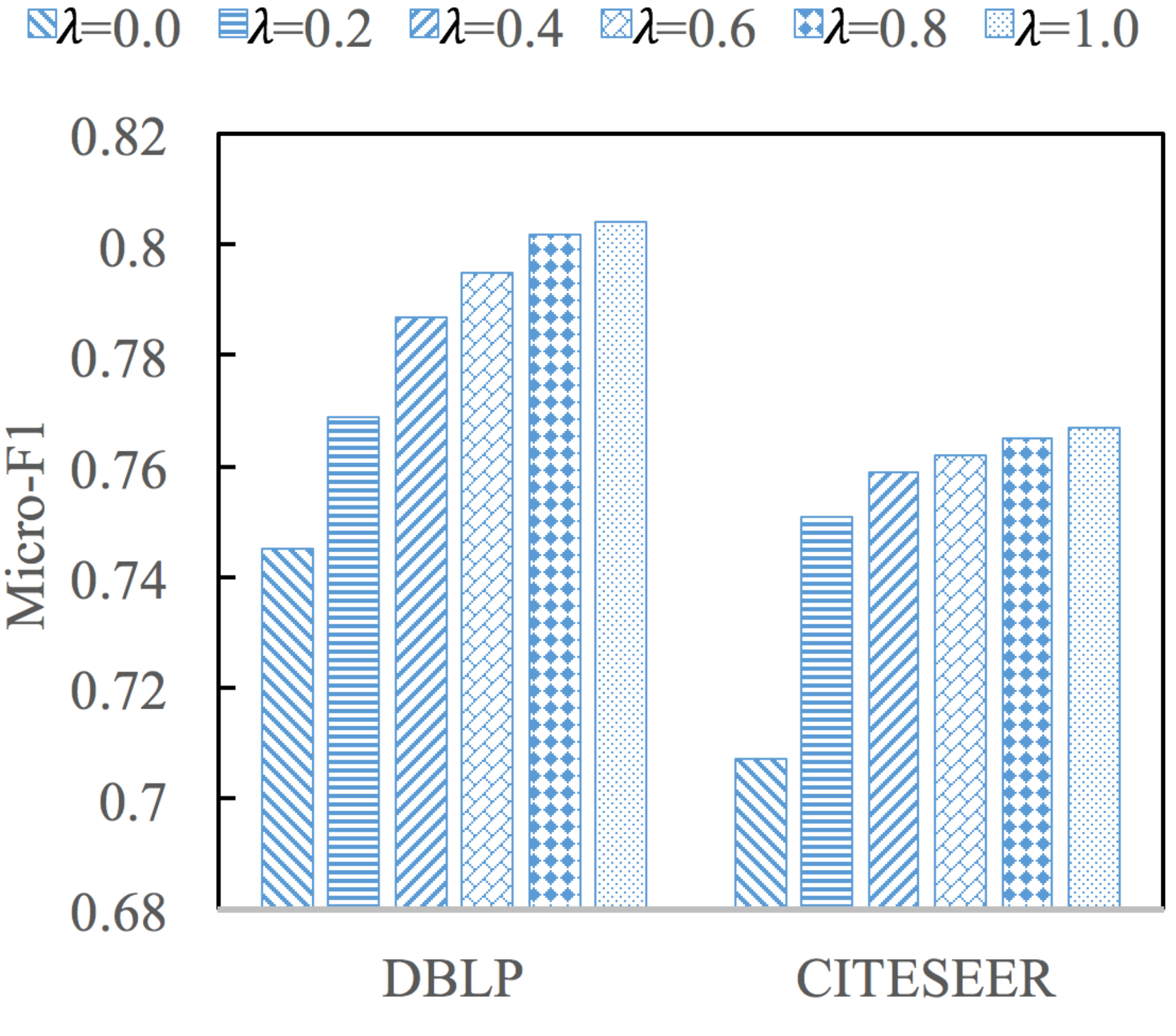}}
	\caption{Performance results with different $\lambda$ (RQ1).}
	\label{Fig:textweight}
\end{figure}

We further explore the impact of $\lambda$ which adjusts the importance of attributes. Both the link prediction task and the node classification task are evaluated under the same evaluation protocols as Section \ref{evalproto}. For a clear comparison, we plot the results in Figure \ref{Fig:textweight}. The link prediction results are reported under training on $80\%$ of links. The node classification results are obtained from training on $50\%$ of labeled nodes.

Due to the fact that $\lambda$ actually can be set to any real number under our learning framework, we first broadly explore the impact of $\lambda$ on the range $[0, 0.01, 0.1, 1, 10, 100]$. Setting $\lambda$ to $0$ returns the pure structure modeling, while setting it to a large number approximates the pure attribute modeling. We found that good results are generally obtained within $[0,1]$ across datasets. When $\lambda$ becomes relatively large and the attribte part overweights the structure part, the performance even becomes worse than pure structure modeling. Therefore, we focus our exploration on the range $[0,1]$ at an interval of $0.2$.

Generally, attributes play an important role in \textit{SNE} as evidenced by the improving performance when $\lambda$ increases. We observe similar trends for both the link prediction and node classification tasks across datasets. If we ignore the attribute information by setting $\lambda = 0.0$, \textit{SNE} degrades to pure structure modeling as detailed in subsection \ref{ss:structure}. Its corresponding performance is the worst for both tasks, as compared to the attributes included counterparts. Moreover, the performance improvements on DBLP and CITESEER are relatively larger. Specifically, we observe a dramatic improvement of performance on CITESEER when $\lambda$ increases from $0.0$ to $0.2$. As there is less link information in these two datasets as shown in Table \ref{Table:datainfo}, the performance improvement indicates that attributes help to alleviate the link sparsity problem.  

In addition, we observe that the pure structure model ($\lambda = 0.0$) outperforms \textit{node2vec} if we further compare the results with Figure \ref{fig:linkprediction} for link prediction and Table \ref{table:nodeclass} for node classification. Since the same $p,q$ setting as \textit{node2vec} are leveraged, we attribute the performance improvements to the non-linearity introduced by the hidden layers.

\subsection{Qualitative Analysis (RQ2)}
\begin{table}[!htp]
	\footnotesize
	\centering
	\renewcommand{\arraystretch}{1.35}
	\caption{Top three results returned by each method (RQ2)}
	\label{table:case}
	\begin{tabular}{l}
		\hline
		\textbf{Query: \textit{Group formation in large social networks: membership,}}\\
		~~~~~~~~~~~~~\textbf{\textit{growth, and evolution}}\footnote{http://dl.acm.org/citation.cfm?id=1150412}\\
		\hline
		\textbf{\textit{SNE}}\\
		1. Structure and evolution of online social networks\\
		2. Discovering temporal communities from social network documents\\
		3. Dynamic social network analysis using latent space models\\
		\hline
		\textbf{\textit{TriDNR}}\\
		1. Influence and correlation in social networks\\
		2. A framework for analysis of dynamic social networks\\
		3. A framework for community identification in dynamic social\\ networks\\
		\hline
		\textbf{\textit{node2vec}}\\
		1. Latent Dirichlet Allocation\\
		2. Maximizing the spread of influence through a social network\\
		3. Mining the network value of customers\\
		\hline
		\textbf{\textit{LINE}}\\
		1. Graphs over time: densification laws, shrinking diameters and\\ possible explanations\\
		2. Maximizing the spread of influence through a social network\\
		3. Relational learning via latent social dimensions\\
		\hline
	\end{tabular}
	\vspace{+0.3cm}
\end{table}
To understand why \textit{SNE} can achieve better results than the other methods, we carry out a case study on the DBLP dataset in this subsection. Given the node representations learned by each method, we retrieve the three most similar papers \textit{w.r.t.} a given query paper. Specifically, we measure the similarity using the cosine distance. For a fair comparison with the structure-based methods, the query paper we choose is a well-cited paper of KDD 2006 named ``\textit{Group formation in large social networks: membership, growth, and evolution}''. According to Google Scholar by 15/1/2017, its citation number reaches 1510. Based on the content of this query paper, we expect that relevant results should be about the structure evolution of groups or communities in social networks. The top results retrieved by different methods are shown in Table~\ref{table:case}.

First of all, we see that \textit{SNE} returns rather relevant results: all the three papers are about dynamic social network analysis and community structures. For example, the first one considers the evolution of structures such as communities in large online social networks. The second result can be viewed as a follow-up work of the query, focusing on discovering temporal communities. While for \textit{TriDNR}, the top result aims to measure social influence between linked individuals but community structures are not of concern.

Regarding methods that only leverage structure information, the results returned by \textit{node2vec} are less similar to the query paper. It seems that \textit{node2vec} tends to find less related but highly cited papers. According to Google Scholar by 15/1/2017, the citation numbers for the first, second and third results are 16908, 4099 and 1815, respectively. This is because the random walk procedure can be easily biased towards the popular nodes that have more links. 
While \textit{SNE} also relies on the walking sequences, it can correct such bias to a certain extent by leveraging attributes. 

Similarly, \textit{LINE} also retrieves less relevant papers. Although the first and second results are related to dynamic social network analysis, all the three results are not concerned with group or community. It might due to the limitations of only modeling first- and second-order proximities while leaving out the abundant attributes.

Based on the above qualitative analysis, we draw the conclusion that using both network structure and attributes benefits the retrieval of similar nodes. 
Compared to the pure structure-based methods, the top returned results of \textit{SNE} are more relevant to the query paper. 
It is worth noting that for this qualitative study, we have purposefully chosen a popular node to migrate the sparsity issue, which actually favors the structure-based methods; even so, the structure-based methods fail at identifying relevant results.
This sheds light on the limitation of solely relying on the network structure for social network embedding, and thus the importance of modeling the rich evidence sources in attributes. 

\subsection{Experiments with Hidden Layers (RQ3)}
\label{ss:rq3}

In this final subsection, we explore the impact of hidden layers on \textit{SNE}. 
It is known that increasing the depth of a neural network can increase the generalization ability for some models~\cite{he2015deep,he2017ncf}, however, it may also degrade the performance due to optimization difficulties~\cite{Glorot10understandingthe}. It is thus curious to see whether using deeper layers can empirically benefit the learning of \textit{SNE}.

\begin{table}[t]
	\centering
	\renewcommand{\arraystretch}{1.35}
	\caption{Performance of link prediction and node classification on DBLP \textit{w.r.t.} different number of hidden layers (RQ3)}
	\label{table:layers}
	\begin{tabular}{l|c|c}
		\hline
		Hidden layers & \textit{AUROC} & \textit{micro-F1}\\
		\hline
		\textit{No Hidden Layers} & 0.9273&0.791\\
		$128$\textit{Softsign}& 0.9418&0.799\\
		$256$\textit{Softsign} $\rightarrow128$\textit{Softsign}& 0.9546&\textbf{0.804}\\
		$512$\textit{Softsign} $\rightarrow256$\textit{Softsign} $\rightarrow 128$\textit{Softsign}~~~~~~&\textbf{0.9589}&0.802\\
		\hline
	\end{tabular}
\end{table}

Table~\ref{table:layers} shows \textit{SNE}'s performance of the link prediction and node classification tasks \textit{w.r.t.} different number of hidden layers on the DBLP dataset. 
The results on other datasets are generally similar, thus we just showcase one here. 
As the size of the last hidden layer determines a \textit{SNE} model's representation ability, we set it to the same number for all models to ensure a fair comparison. 
Note that for each setting (row), we have re-tuned the hyper-parameters to fully exploit the model's performance. 

First, we can see the trend that with more hidden layers, the performance is improved. 
This indicates the positive effect of using a deeper architecture for \textit{SNE}, which indeed increases its generalization ability and boost its performance. 
The trade-off, however, is the server CPU time needed for the training procedure. 
Specifically, on our modest commodity server (Intel Xeon CPU
of 2.40GHz), a one-layer \textit{SNE} takes 25.6 seconds, while a three-layer \textit{SNE} takes 81.9 seconds for one epoch. 
We stopped exploring deeper models, as the current \textit{SNE} uses fully connected layers, which become difficult to optimize and can be easily over-fitting and degrading with more layers~\cite{Glorot10understandingthe}. 
The diminishing improvement of results in Table~\ref{table:layers} also implies the potential problem. 
To address it, modern neural network designs shall be applied, such as the residual units and highway networks~\cite{he2015deep}. We leave this possibility for future work.

It is worth noting that when there is no hidden layer, \textit{SNE}'s performance is rather weak, which is in the same level as \textit{TriDNR}.
With one more layer, the performance is significantly improved. 
This demonstrates the usefulness of learning structure--attribute interactions in a non-linear way. To justify this, we have further tried to replace the softsign activation function with the identity function, \textit{i.e.,} using a linear function above the concatenation of structure and attribute embedding vectors. However, the performance is much worse than that of using the non-linear softsign function.

\section{Conclusion}
To learn informative representations for social network data, it is crucial to account for both network structure and attribute information. To this end, we proposed a generic framework for embedding social networks by capturing both the
\textit{structural proximity} and \textit{attribute proximity}.
We adopted a deep neural network architecture to model the complex interrelations between structural information and attributes. Extensive experiments show that \textit{SNE} can learn informative representations for social networks and achieve superior performance on the tasks of link prediction and node classification comparing to other representation learning methods.

This work has tackled representation learning on social networks by leveraging both structural and attribute information. While social networks are rich sources of information containing more than links and textual attributes, we will study the following directions in future. First, we will enhance our SNE framework by fusing data from multiple modalities. It is reported that over 45\% tweets contain images in Weibo \cite{chen2016context}, making it urgent and meaningful to perform network embedding with multi-modal data \cite{zhang2017regions}. Second, we will develop (semi-)supervised variant for SNE, so as to learning task-oriented embeddings to tailor for a specific task. Third, we are interested in exploring how to capture the evolution nature of social networks, such as new users and social relations by using temporal-aware recurrent neural networks. Lastly, we will consider improving the efficiency of SNE by learning to hash techniques \cite{zhang2016discrete} to make it suitable for large-scale industrial use.

\ifCLASSOPTIONcompsoc
  \section*{Acknowledgments}
\else
  \section*{Acknowledgment}
\fi

This research is supported by the NExT research center, which is supported by the National Research Foundation, Prime Minister's Office, Singapore under its IRC@SG Funding Initiative. We warmly thank all the anonymous reviewers for their time and efforts.



\begin{thebibliography}{10}
	\providecommand{\url}[1]{#1}
	\csname url@samestyle\endcsname
	\providecommand{\newblock}{\relax}
	\providecommand{\bibinfo}[2]{#2}
	\providecommand{\BIBentrySTDinterwordspacing}{\spaceskip=0pt\relax}
	\providecommand{\BIBentryALTinterwordstretchfactor}{4}
	\providecommand{\BIBentryALTinterwordspacing}{\spaceskip=\fontdimen2\font plus
		\BIBentryALTinterwordstretchfactor\fontdimen3\font minus
		\fontdimen4\font\relax}
	\providecommand{\BIBforeignlanguage}[2]{{%
			\expandafter\ifx\csname l@#1\endcsname\relax
			\typeout{** WARNING: IEEEtran.bst: No hyphenation pattern has been}%
			\typeout{** loaded for the language `#1'. Using the pattern for}%
			\typeout{** the default language instead.}%
			\else
			\language=\csname l@#1\endcsname
			\fi
			#2}}
	\providecommand{\BIBdecl}{\relax}
	\BIBdecl
	
	\bibitem{wang2017}
	X.~Wang, L.~Nie, X.~Song, D.~Zhang, and T.-S. Chua, ``Unifying virtual and
	physical worlds: Learning toward local and global consistency,'' \emph{ACM
		Transactions on Information Systems}, vol.~36, no.~1, p.~4, 2017.
	
	\bibitem{he2017birank}
	X.~He, M.~Gao, M.-Y. Kan, and D.~Wang, ``Birank: Towards ranking on bipartite
	graphs,'' \emph{IEEE Transactions on Knowledge and Data Engineering},
	vol.~29, no.~1, pp. 57--71, 2017.
	
	\bibitem{perozzi2014deepwalk}
	B.~Perozzi, R.~Al-Rfou, and S.~Skiena, ``Deepwalk: Online learning of social
	representations,'' in \emph{SIGKDD}, 2014, pp. 701--710.
	
	\bibitem{chang2015heterogeneous}
	S.~Chang, W.~Han, J.~Tang, G.-J. Qi, C.~C. Aggarwal, and T.~S. Huang,
	``Heterogeneous network embedding via deep architectures,'' in \emph{SIGKDD},
	2015, pp. 119--128.
	
	\bibitem{grovernode2vec}
	A.~Grover and J.~Leskovec, ``node2vec: Scalable feature learning for
	networks,'' in \emph{SIGKDD}, 2016, pp. 855--864.
	
	\bibitem{tang2015line}
	J.~Tang, M.~Qu, M.~Wang, M.~Zhang, J.~Yan, and Q.~Mei, ``Line: Large-scale
	information network embedding,'' in \emph{WWW}, 2015, pp. 1067--1077.
	
	\bibitem{burger2011discriminating}
	J.~D. Burger, J.~Henderson, G.~Kim, and G.~Zarrella, ``Discriminating gender on
	twitter,'' in \emph{EMNLP}, 2011, pp. 1301--1309.
	
	\bibitem{pennacchiotti2011democrats}
	M.~Pennacchiotti and A.-M. Popescu, ``Democrats, republicans and starbucks
	afficionados: user classification in twitter,'' in \emph{SIGKDD}, 2011, pp.
	430--438.
	
	\bibitem{traud2012social}
	A.~L. Traud, P.~J. Mucha, and M.~A. Porter, ``Social structure of facebook
	networks,'' \emph{Physica A: Statistical Mechanics and its Applications}, pp.
	4165--4180, 2012.
	
	\bibitem{wangstructural}
	D.~Wang, P.~Cui, and W.~Zhu, ``Structural deep network embedding,'' in
	\emph{SIGKDD}, 2016, pp. 1225--1234.
	
	\bibitem{robins2011exponential}
	G.~Robins, ``Exponential random graph models for social networks,''
	\emph{Encyclopaedia of Complexity and System Science, Springer}, 2011.
	
	\bibitem{lazarsfeld1954friendship}
	P.~F. Lazarsfeld, R.~K. Merton \emph{et~al.}, ``Friendship as a social process:
	A substantive and methodological analysis,'' \emph{Freedom and control in
		modern society}, pp. 18--66, 1954.
	
	\bibitem{laumann1966prestige}
	E.~O. Laumann, \emph{Prestige and association in an urban community: An
		analysis of an urban stratification system}.\hskip 1em plus 0.5em minus
	0.4em\relax Bobbs-Merrill Company, 1966.
	
	\bibitem{mcpherson2001birds}
	M.~McPherson, L.~Smith-Lovin, and J.~M. Cook, ``Birds of a feather: Homophily
	in social networks,'' \emph{Annual review of sociology}, pp. 415--444, 2001.
	
	\bibitem{kurth1970friendships}
	S.~B. Kurth, ``Friendships and friendly relations,'' \emph{Social
		relationships}, pp. 136--170, 1970.
	
	\bibitem{mcpherson1987homophily}
	J.~M. McPherson and L.~Smith-Lovin, ``Homophily in voluntary organizations:
	Status distance and the composition of face-to-face groups,'' \emph{American
		sociological review}, pp. 370--379, 1987.
	
	\bibitem{fiore2005homophily}
	A.~T. Fiore and J.~S. Donath, ``Homophily in online dating: when do you like
	someone like yourself?'' in \emph{CHI '05}, 2005, pp. 1371--1374.
	
	\bibitem{kossinets2009origins}
	G.~Kossinets and D.~J. Watts, ``Origins of homophily in an evolving social
	network1,'' \emph{American journal of sociology}, pp. 405--450, 2009.
	
	\bibitem{roweis2000nonlinear}
	S.~T. Roweis and L.~K. Saul, ``Nonlinear dimensionality reduction by locally
	linear embedding,'' \emph{Science}, pp. 2323--2326, 2000.
	
	\bibitem{tenenbaum2000global}
	J.~B. Tenenbaum, V.~De~Silva, and J.~C. Langford, ``A global geometric
	framework for nonlinear dimensionality reduction,'' \emph{science}, pp.
	2319--2323, 2000.
	
	\bibitem{belkin2001laplacian}
	M.~Belkin and P.~Niyogi, ``Laplacian eigenmaps and spectral techniques for
	embedding and clustering,'' in \emph{NIPS}, 2001, pp. 585--591.
	
	\bibitem{huang2017label}
	X.~Huang, J.~Li, and X.~Hu, ``Label informed attributed network embedding,'' in
	\emph{WSDM}, 2017.
	
	\bibitem{yang2015network}
	C.~Yang, Z.~Liu, D.~Zhao, M.~Sun, and E.~Y. Chang, ``Network representation
	learning with rich text information,'' in \emph{IJCAI}, 2015, pp. 2111--2117.
	
	\bibitem{shirui2016}
	S.~Pan, J.~Wu, X.~Zhu, C.~Zhang, and Y.~Wang, ``Tri-party deep network
	representation,'' in \emph{IJCAI}, 2016, pp. 1895--1901.
	
	\bibitem{le2014distributed}
	Q.~V. Le and T.~Mikolov, ``Distributed representations of sentences and
	documents,'' in \emph{ICML}, 2014, pp. 1188--1196.
	
	\bibitem{cheng2016wide}
	H.-T. Cheng, L.~Koc, J.~Harmsen, T.~Shaked, T.~Chandra, H.~Aradhye,
	G.~Anderson, G.~Corrado, W.~Chai, M.~Ispir \emph{et~al.}, ``Wide \& deep
	learning for recommender systems,'' in \emph{Workshop on DLRS}, 2016, pp.
	7--10.
	
	\bibitem{weston2012deep}
	J.~Weston, F.~Ratle, H.~Mobahi, and R.~Collobert, ``Deep learning via
	semi-supervised embedding,'' in \emph{Neural Networks: Tricks of the Trade},
	2012, pp. 639--655.
	
	\bibitem{yang2016icml}
	Z.~Yang, W.~Cohen, and R.~Salakhudinov, ``Revisiting semi-supervised learning
	with graph embeddings,'' in \emph{ICML}, 2016, pp. 40--48.
	
	\bibitem{cao2015grarep}
	S.~Cao, W.~Lu, and Q.~Xu, ``Grarep: Learning graph representations with global
	structural information,'' in \emph{CIKM}, 2015, pp. 891--900.
	
	\bibitem{he2016fast}
	X.~He, H.~Zhang, M.-Y. Kan, and T.-S. Chua, ``Fast matrix factorization for
	online recommendation with implicit feedback,'' in \emph{SIGIR}, 2016, pp.
	549--558.
	
	\bibitem{mikolov2013efficient}
	T.~Mikolov, I.~Sutskever, K.~Chen, G.~S. Corrado, and J.~Dean, ``Distributed
	representations of words and phrases and their compositionality,'' in
	\emph{NIPS}, 2013, pp. 3111--3119.
	
	\bibitem{he2017ncf}
	X.~He, L.~Liao, H.~Zhang, L.~Nie, X.~Hu, and T.~Chua, ``Neural collaborative
	filtering,'' in \emph{WWW}, 2017.
	
	\bibitem{luo2011cauchy}
	D.~Luo, F.~Nie, H.~Huang, and C.~H. Ding, ``Cauchy graph embedding,'' in
	\emph{ICML}, 2011, pp. 553--560.
	
	\bibitem{pennington2014glove}
	J.~Pennington, R.~Socher, and C.~D. Manning, ``Glove: Global vectors for word
	representation,'' in \emph{EMNLP}, 2014, pp. 1532--1543.
	
	\bibitem{TACL570}
	O.~Levy, Y.~Goldberg, and I.~Dagan, ``Improving distributional similarity with
	lessons learned from word embeddings,'' \emph{Transactions of the Association
		for Computational Linguistics}, 2015.
	
	\bibitem{FM}
	S.~Rendle, ``Factorization machines,'' in \emph{ICDM}, 2010, pp. 995--1000.
	
	\bibitem{shan2016deep}
	Y.~Shan, T.~R. Hoens, J.~Jiao, H.~Wang, D.~Yu, and J.~Mao, ``Deep crossing:
	Web-scale modeling without manually crafted combinatorial features,'' in
	\emph{SIGKDD}, 2016, pp. 255--262.
	
	\bibitem{he2017neu}
	X.~He and T.-S. Chua, ``Neural factorization machines,'' in \emph{SIGIR}, 2017,
	p. to appear.
	
	\bibitem{he2015deep}
	K.~He, X.~Zhang, S.~Ren, and J.~Sun, ``Deep residual learning for image
	recognition,'' in \emph{CVPR}, 2016, pp. 770--778.
	
	\bibitem{wang2017your}
	S.~Wang, Y.~Wang, J.~Tang, K.~Shu, S.~Ranganath, and H.~Liu, ``What your images
	reveal: Exploiting visual contents for point-of-interest recommendation,'' in
	\emph{WWW}, 2017, pp. 391--400.
	
	\bibitem{kingma2014adam}
	D.~Kingma and J.~Ba, ``Adam: A method for stochastic optimization,'' in
	\emph{ICLR}, 2015, pp. 1--15.
	
	\bibitem{duchi2011adaptive}
	J.~Duchi, E.~Hazan, and Y.~Singer, ``Adaptive subgradient methods for online
	learning and stochastic optimization,'' \emph{Journal of Machine Learning
		Research}, pp. 2121--2159, 2011.
	
	\bibitem{tieleman2012lecture}
	T.~Tieleman and G.~Hinton, ``Lecture 6.5-rmsprop, coursera: Neural networks for
	machine learning,'' Tech. Rep., 2012.
	
	\bibitem{IoffeS15}
	S.~Ioffe and C.~Szegedy, ``Batch normalization: Accelerating deep network
	training by reducing internal covariate shift,'' in \emph{ICML}, 2015, pp.
	448--456.
	
	\bibitem{koren2008factorization}
	Y.~Koren, ``Factorization meets the neighborhood: a multifaceted collaborative
	filtering model,'' in \emph{SIGKDD}, 2008, pp. 426--434.
	
	\bibitem{wasserman1994social}
	S.~Wasserman and K.~Faust, \emph{Social network analysis: Methods and
		applications}, 1994.
	
	\bibitem{tang2008}
	J.~Tang, J.~Zhang, L.~Yao, J.~Li, L.~Zhang, and Z.~Su, ``Arnetminer: extraction
	and mining of academic social networks,'' in \emph{SIGKDD}, 2008, pp.
	990--998.
	
	\bibitem{zou2007receiver}
	K.~H. Zou, A.~J. O’Malley, and L.~Mauri, ``Receiver-operating characteristic
	analysis for evaluating diagnostic tests and predictive models,''
	\emph{Circulation}, pp. 654--657, 2007.
	
	\bibitem{fan2008liblinear}
	R.-E. Fan, K.-W. Chang, C.-J. Hsieh, X.-R. Wang, and C.-J. Lin, ``Liblinear: A
	library for large linear classification,'' \emph{Journal of machine learning
		research}, pp. 1871--1874, 2008.
	
	\bibitem{Glorot10understandingthe}
	X.~Glorot and Y.~Bengio, ``Understanding the difficulty of training deep
	feedforward neural networks,'' in \emph{AISTATS}, 2010, pp. 249--256.
	
	\bibitem{chen2016context}
	T.~Chen, X.~He, and M.-Y. Kan, ``Context-aware image tweet modelling and
	recommendation,'' in \emph{MM}, 2016, pp. 1018--1027.
	
	\bibitem{zhang2017regions}
	C.~Zhang, K.~Zhang, Q.~Yuan, H.~Peng, Y.~Zheng, T.~Hanratty, S.~Wang, and
	J.~Han, ``Regions, periods, activities: Uncovering urban dynamics via
	cross-modal representation learning,'' in \emph{WWW}, 2017, pp. 361--370.
	
	\bibitem{zhang2016discrete}
	H.~Zhang, F.~Shen, W.~Liu, X.~He, H.~Luan, and T.-S. Chua, ``Discrete
	collaborative filtering,'' in \emph{SIGIR}, 2016, pp. 325--334.
	
\end{thebibliography}
\end{document}